\documentclass[longauth]{aa}
\usepackage[varg]{txfonts}
\usepackage{siunitx}
\usepackage{amsmath}
\usepackage{upgreek}
\usepackage{graphicx}
\usepackage{bbold}
\usepackage{color,soul}
\usepackage[colorlinks=true,citecolor=blue,linkcolor=blue,urlcolor=blue]{hyperref}
\usepackage{float}
\usepackage{placeins}
\usepackage{subcaption}
\usepackage{xcolor}
\usepackage{multicol}
\usepackage[labelfont=bf]{caption}
\usepackage{ulem}
\newcommand{\degree}{\mbox{$^{\circ}$}}
\bibpunct{(}{)}{;}{a}{}{,}

\DeclareSIUnit\mag{mag}
\begin{document} 
\title{First Light for GRAVITY Wide:\\Large Separation Fringe Tracking for the Very Large Telescope Interferometer}
\titlerunning{First Light for GRAVITY Wide}
\authorrunning{GRAVITY+ Collaboration}

% \usepackage{natbib}
% \bibpunct{(}{)}{;}{a}{}{,}

\author{GRAVITY+ Collaboration\thanks{GRAVITY+ is developed by the Max Planck Institute for extraterrestrial Physics, the Institute National des Sciences de l'Univers du CNRS (INSU) with its institutes LESIA / Paris Observatory-PSL, IPAG / Grenoble Observatory, Lagrange / Côte d’Azur Observatory and CRAL / Lyon Observatory, the Max Planck Institute for Astronomy, the University of Cologne, the CENTRA - Centro de Astrofisica e Gravita\c c\~ao, the University of Southampton, the Katholieke Universiteit Leuven and the European Southern Observatory. \newline%$\,\,\,\,\,\,$ 
Corresponding authors: A.~Drescher (email: drescher$@$mpe.mpg.de) and J.~Woillez (email: jwoillez$@$eso.org)
    }: 
R.~Abuter\inst{8}
\and F.~Allouche\inst{23}
\and A.~Amorim\inst{6,12}
\and C.~Bailet\inst{23}
\and M.~Baub\"ock\inst{15,1}
\and J.-P.~Berger\inst{5}
\and P.~Berio\inst{23}
\and A.~Bigioli\inst{19}
% ...
}

\institute{
Max Planck Institute for extraterrestrial Physics, Giessenbachstra{\ss}e~1, 85748 Garching, Germany
\and LESIA, Observatoire de Paris, Universit\'e PSL, Sorbonne Universit\'e, Universit\'e Paris Cit\'e, CNRS, 5 place Jules Janssen, 92195 Meudon, France
\and Max Planck Institute for Astronomy, K\"onigstuhl 17, 69117 Heidelberg, Germany
\and $1^{\rm st}$ Institute of Physics, University of Cologne, Z\"ulpicher Stra{\ss}e 77, 50937 Cologne, Germany
\and Univ. Grenoble Alpes, CNRS, IPAG, 38000 Grenoble, France
\and Universidade de Lisboa - Faculdade de Ci\^encias, Campo Grande, 1749-016 Lisboa, Portugal 
\and Faculdade de Engenharia, Universidade do Porto, rua Dr. Roberto Frias, 4200-465 Porto, Portugal 
\and European Southern Observatory, Karl-Schwarzschild-Stra{\ss}e 2, 85748 Garching, Germany
\and European Southern Observatory, Casilla 19001, Santiago 19, Chile
\and Sterrewacht Leiden, Leiden University, Postbus 9513, 2300 RA Leiden, The Netherlands
\and Departments of Physics and Astronomy, Le Conte Hall, University of California, Berkeley, CA 94720, USA
\and CENTRA - Centro de Astrof\'{\i}sica e Gravita\c c\~ao, IST, Universidade de Lisboa, 1049-001 Lisboa, Portugal
\and Department of Physics, Technical University Munich, James-Franck-Straße 1,  85748 Garching, Germany
\and Max Planck Institute for Radio Astronomy, Auf dem H\"ugel 69, 53121 Bonn, Germany
\and Department of Physics, University of Illinois, 1110 West Green Street, Urbana, IL 61801, USA
\and Hamburger Sternwarte, Universität Hamburg, Gojenbergsweg 112, 21029 Hamburg, Germany
\and School of Physics \& Astronomy, University of Southampton, Southampton, SO17 1BJ, UK
\and Max Planck Institute for Astrophysics, Karl-Schwarzschild-Str. 1, 85741 Garching, Germany
\and Institute of Astronomy, KU Leuven, Celestijnenlaan 200D, B-3001, Leuven, Belgium
\and European Space Agency, European Space Astronomy Centre, Madrid, Spain
\and Institute of Astronomy, Madingley Road, Cambridge CB3 0HA, UK
\and Research School of Astronomy and Astrophysics, College of Science, Australian National University, Canberra, Australia
\and Universit\'e Côte d'Azur, Observatoire de la C\^ote d'Azur, CNRS, Laboratoire Lagrange, France
% ...
}

\date{Received May 4, 2022; accepted May 30, 2022}

% \abstract{}{}{}{}{} 
% 5 {} token are mandatory
%%%%%%%%%%%%%%%%%%%%%%%%%
% ABSTRACT
%%%%%%%%%%%%%%%%%%%%%%%%

\abstract{
GRAVITY+ is the upgrade of GRAVITY and the Very Large Telescope Interferometer (VLTI) with wide-separation fringe tracking, new adaptive optics, and laser guide stars on all four 8~m Unit Telescopes (UTs), for ever fainter, all-sky, high contrast, milliarcsecond interferometry. Here we present the design and first results of the first phase of GRAVITY+, called GRAVITY Wide. GRAVITY Wide combines the dual-beam capabilities of the VLTI and the GRAVITY instrument to increase the maximum separation between the science target and the reference star from 2 arcseconds with the 8 m UTs up to several 10 arcseconds, limited only by the Earth's turbulent atmosphere. This increases the sky-coverage of GRAVITY by two orders of magnitude, opening up milliarcsecond resolution observations of faint objects, and in particular the extragalactic sky. The first observations in 2019 -- 2022 include first infrared interferometry of two redshift $z\sim2$ quasars, interferometric imaging on the binary system HD 105913A, and repeated observations of multiple star systems in the Orion Trapezium Cluster. We find the coherence loss between the science object and fringe-tracking reference star well described by the turbulence of the Earth's atmosphere. We confirm that the larger apertures of the UTs result in higher visibilities for a given separation due to larger overlap of the projected pupils on sky and give predictions for visibility loss as a function of separation to be used for future planning.}



%\begin{document} 
\maketitle
\section{Introduction} 
\label{sec:introduction}
\citet{1992A&A...262..353S} describe for the first time an optical interferometer observing simultaneously two widely separated targets contained inside the atmospheric turbulence isopistonic patch. The Palomar Testbed Interferometer (PTI) \citep{1999ApJ...510..505C} represents the first implementation of this dual-field technique, where star separators are located at the focus of the telescope and deliver two independent beams allowing to operate simultaneously two interferometric instuments. At the time however, the emphasis was primarily on astrometry, i.e. measuring the angular distance between the two targets, in preparation for NASA's space astrometry missions for exoplanets. The possibility to use the technique to observe much fainter targets was tentatively explored and presented in \citet{2003AJ....125.1623L}, but remained within the limiting magnitudes of the PTI around $m_K=5$. On the VLTI \citep{1990Msngr..60....1B}, the dual-field instrument PRIMA \citep{2008NewAR..52..199D} was foreseen to deliver astrometric and phase-referencing capabilities. The emphasis remained on astrometry rather than pushing the sensitivity of the interferometer, until the project was discontinued, in face of the competition with Gaia \citep{2001A&A...369..339P}, while preserving the dual-field capability of the infrastructure. The first dual-field phase-referenced observations to demonstrate a sensitivity improvement were carried out by the ASTRA instrument \citep{2014ApJ...783..104W} of the Keck Interferometer \citep{2013PASP..125.1226C}, reaching a magnitude of m$_K$ = 12.5, which was about 10 times fainter than contemporaneous direct observations. The scientific exploitation of this nascent capability was however cut short by the early demise of this facility in July 2012.

The sensitivity revolution was finally delivered by the GRAVITY instrument \citep{2017A&A...602A..94G} at the VLTI. It has transformed high angular resolution astronomy with the first interferometric instrument to routinely offer milliarcsecond (mas) resolution imaging for objects as faint as $m_K=19-20$, a sensitivity increase by more than a factor thousand over previous interferometers, 30-100 microarcsecond ($\mu$as) astrometry, and microarcsecond differential spectro-astrometry. The key to success are technical breakthroughs on several fronts, including the development of quasi-noiseless infrared detectors \citep{2019SPIE11180E..6LF}, infrared single-mode waveguides and integrated optics \citep{2018A&A...614A..70P}, robust fringe tracking \citep{2019A&A...624A..99L}, infrared adaptive optics \citep{2016SPIE.9909E..2LS}, and laser metrology \citep{2012SPIE.8445E..1OG}, as well as performance improvements all over the VLTI observatory \citep{2018SPIE10701E..03W}. GRAVITY is also the first interferometer to routinely offer dual-field interferometry, for which a bright reference star is used to stabilize and phase-reference the interferogram of the science object.\looseness=-2

In the first five years of science operation, GRAVITY brought groundbreaking results covering a broad range of astrophysical science: It has provided the strongest experimental evidence that the compact mass in the Galactic Center (Sgr A*) is a black hole, including the first detection of the gravitational redshift \citep{2018A&A...615L..15G} and the Schwarzschild precession \citep{2020A&A...636L...5G} in the orbit of the star S2 around the black hole. Further, GRAVITY has detected orbital motion of hot gas close to the innermost stable orbit of the black hole \citep{2018A&A...618L..10G}, and performed the most precise measurement of the black hole's mass and distance \citep{2019A&A...625L..10G}, surrounding mass distribution \citep{2022A&A...657L..12G}, and tests of the Einstein equivalence principle \citep{2019PhRvL.122j1102A}. GRAVITY has provided high resolution spectra of the atmosphere of several exoplanets, including HR8799e \citep{2019A&A...623L..11G} and $\beta$ Pic b \citep{2020A&A...633A.110G}. On the latter, the measured C/O ratio indicates that this planet has undergone substantial core accretion and planetesimal enrichment. Additionally, GRAVITY has delivered the first direct detection of a radial velocity planet $\beta$ Pic c \citep{2020A&A...642L...2N}, and the measurement of the mass of an exoplanet from the astrometry of a second planet \citep{2021A&A...654L...2L}. GRAVITY was also the first instrument to spatially resolve a quasar broad line region (BLR) \citep{2018Natur.563..657G}, and to image at milliarcsecond resolution the dust sublimation region around a Seyfert 2 active galactic nucleus (AGN) \citep{2020A&A...634A...1G}. It has also provided a comprehensive dataset of spatially resolved disks of young stellar objects \citep{2019A&A...632A..53G, 2021A&A...648A..37G, 2021A&A...655A.112G, 2021A&A...645A..50G, 2021A&A...655A..73G,  2021A&A...654A..97G}, and spatially resolved the magnetospheric accretion onto a T Tauri star \citep{2020Natur.584..547G}. Further, GRAVITY resolved for the first time the two images produced by gravitational microlenses \citep{2019ApJ...871...70D}.

Until now, the number of observable targets with the dual-feed mode is limited by the requirement that the fringe-tracking (FT) source and the science target (SC) have to be within the field of view of the VLTI, which is 2 arcseconds (arcsec) in diameter for the Unit Telescopes (UTs) and 4 arcsec in diameter for the Auxiliary Telescopes (ATs). This requirement can be overcome by implementing wide-angle off-axis fringe tracking, where we enlarge the separation between FT and SC up to about 30 arcsec, limited by the atmospheric turbulence. To break the limitation in separation between the two fields of GRAVITY, they are separated at the telescope level and finally overlapped at the entrance of GRAVITY. This implementation, which we refer to as GRAVITY Wide, is one of the primary components of the ongoing upgrades to the VLTI and GRAVITY. This improved instrument is called GRAVITY+ \citep{white_paper_gravityplus}. Besides GRAVITY Wide, GRAVITY+ also includes the implementation of new deformable mirrors and state-of-the-art adaptive optics (AO) wavefront sensors, an improved instrument throughput and vibration control, as well as laser guide stars on all four 8 m UTs. These upgrades will enable fringe tracking on objects as faint as $m_K=13$ and, together with GRAVITY Wide, enable all-sky interferometry with high resolution imaging at milliarcsec accuracy with a limiting magnitude of $m_K=22$.

The increased FT--SC angular separation will make it possible to observe faint targets with fringe tracking on a nearby bright source that can be picked from a much larger area. This will open up observations and discoveries in different areas of astronomy, such as: to spatially resolve young stellar objects in their embedded phase, to study the multiplicity of massive stars in the Small and Large Magellanic Cloud, to constrain for the first time intermediate mass black holes with accurate motions of stars in globular clusters, to discover single stellar-mass black holes and free floating planets via microlensing, and to probe supermassive black holes in active galactic nuclei out to beyond $z \approx 2$ \citep{white_paper_gravityplus} and in nearby inactive galaxies with transient tidal disruption events.\looseness=-2

The enlargement of the FT--SC separation, however, brings along a challenge. In ground-based optical and infrared interferometry, atmospheric turbulence plays an important role \citep{1966JOSA...56.1372F}. Local changes in the temperature and humidity in the atmosphere lead to changes in the refractive index of air. When an initially flat wavefront from a distant science object enters the atmosphere, it gets distorted. Adaptive optics and the fringe tracker are able to correct these distortions. However, when the separation between SC and FT increases, the correction degrades on the SC due to residual wavefront errors in the direction of the SC. This effect is called anisoplanatism \citep{Fried:82} and it becomes more severe for larger off-axis separations. Thus, it is important to understand the behaviour of the atmosphere and investigate the influence of atmospheric turbulence on large separation fringe tracking.

\hfill \break
In this paper, we present the first wide-angle interferometric observations with four telescopes, performed with GRAVITY Wide. We describe the instrumental changes of GRAVITY to GRAVITY Wide in Sect.~\ref{sec:gwidemode}. Sect.~\ref{sec:observations}, we present the GRAVITY Wide data, and demonstrate first GRAVITY Wide science. From observations of multiple star systems in the Orion Trapezium Cluster, we derive a new orbit for $\theta^1$ Ori B, and refine the orbits of $\theta^1$ Ori C and $\theta^1$ Ori D. We present the separation and flux ratio of the binary HD 105913A, and the successful detection of fringes across the H$\alpha$ line for two $z>2$ quasars, SDSS J161513.84+084914.4 ($z=2.33$, hereafter SDSS1615) and LAMOST J092034.16+065717.9 ($z=2.46$, hereafter LAMOST09). In Sect.~\ref{sec:discussion}, we discuss the influence of atmospheric turbulence on the new large separation fringe tracking mode with 17 observed FT--SC pairs with separations up to 32 arcsec. Finally, we give a summary and an outlook for future GRAVITY+ science in Sect.~\ref{sec:summary}.

\begin{figure}[t]
\centering
\includegraphics[width=0.45\textwidth]{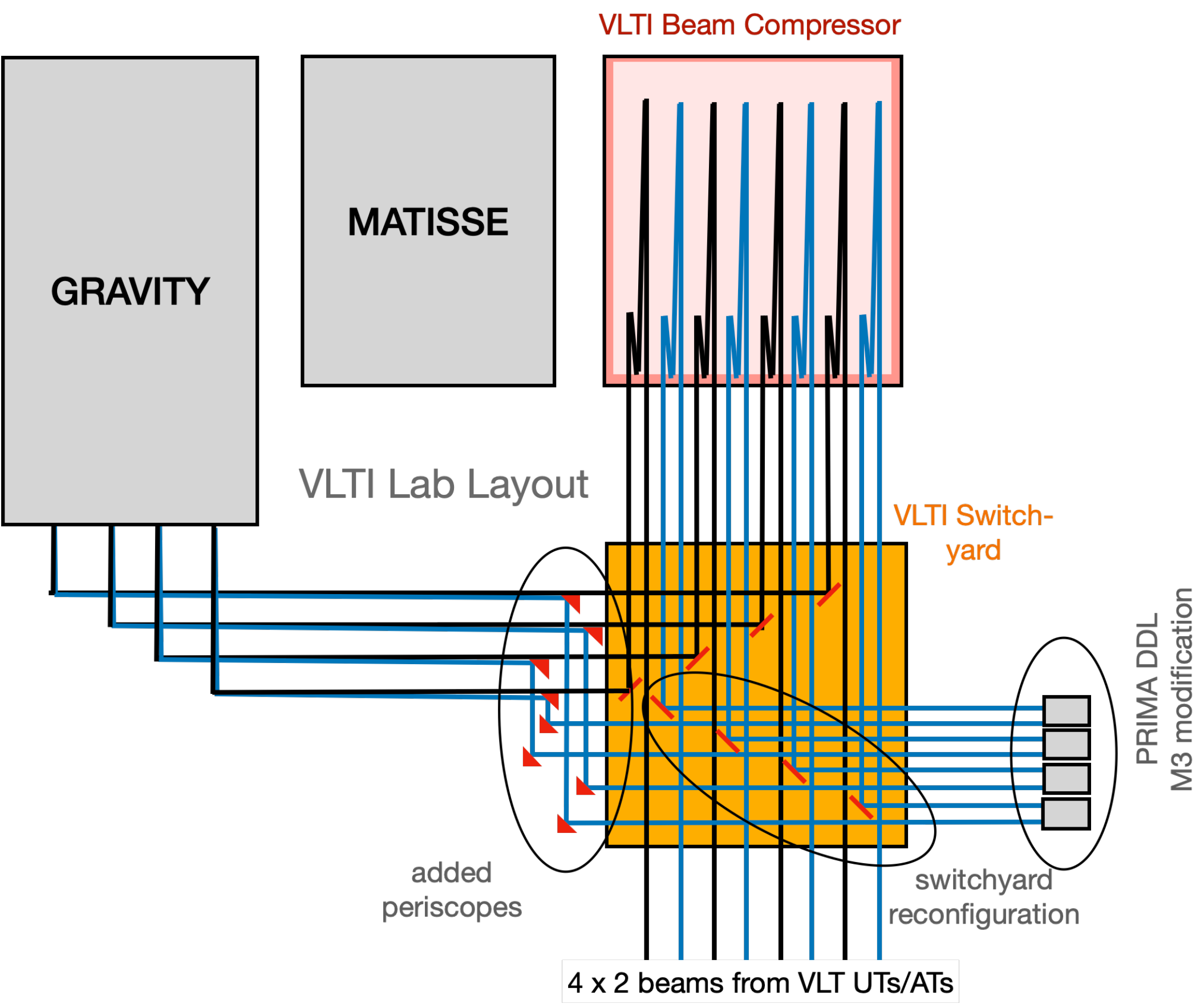}
\caption{Modifications made to the VLTI switchyard in December 2021 to implement the first phase of GRAVITY Wide. Located in the VLTI laboratory underneath the VLTI platform, the switchyard receives the light from the main VLTI delay lines and directs the light to the various downstream instruments such as GRAVITY, MATISSE, and PIONIER. Typically, the light first enters the beam compressors to convert the beam diameters from 80\,mm down to 18\,mm. To allow for simultaneous observations of FT targets separated by more than 2 arcsec from the SC we enabled the use of the B beams from the VLTI. We added a four-fold periscope consisting of eight flat fold mirrors to merge the A (black) and B (blue) beams with a 2 arcsec separation. For this it was also necessary to rearrange four of the eight main switchyard mirrors, and reactivated the original PRIMA differential delay lines.}
\label{fig:gwide_switchyard}
\end{figure}

\section{The GRAVITY Wide upgrade} 
\label{sec:gwidemode}
\subsection{Instrumental changes}
The original design of the VLTI incorporates dual-field interferometric capability. Two subsections of the telescopes fields of view, each about 2 arcsec wide (4 arcsec for the ATs) and separated by up to 60 arcsec are picked up by the star separators (STS) located at the coud\'e focus of each telescope and propagated through the VLTI delay lines into the VLTI laboratory. The two beams are generally referred to as the A and B beams, where commonly only the A beams are used in non-dual-field applications.

\begin{figure}[t]
\centering
\includegraphics[width=0.45\textwidth]{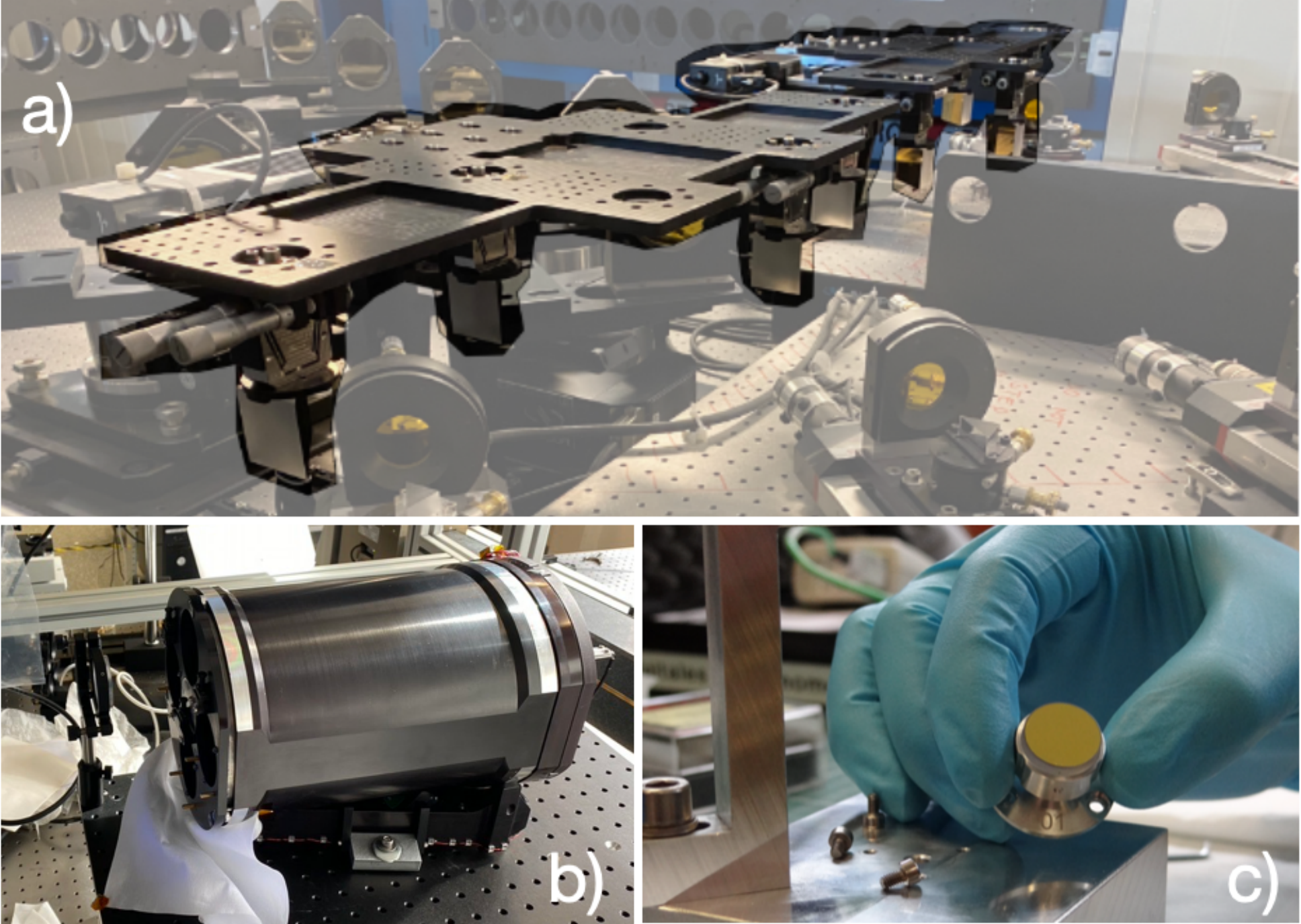}
\caption{Photographs of the added or modified hardware for the implementation of GRAVITY Wide. \textit{Panel a)} The newly added periscopes merge the A and B beams of the respective four UT or ATs.  Due to space constraints they were mounted hanging down from two motorized bridge structures. \textit{Panel b)} To compensate for the differential optical path length between the A and B-beams, we now use the differential delay lines formerly belonging to PRIMA. \textit{Panel c)} These DDLs, however, needed to be modified to relay the pupil to the appropriate location for GRAVITY. The PRIMA DDLs consist of a three mirror cat's eye with five optical reflections. We replaced the tertiary mirrors to adjust the pupil relay. Our beam path design allowed us to choose identical radii of curvature for all tertiary mirrors.}
\label{fig:gwide_ddl}
\end{figure}

To enable wide-mode observations we introduced new optics that merge the A and B beams on the VLTI switch yard before feeding them to GRAVITY. This is achieved through four periscopes that pick up the B beams and translate them laterally to bring them into overlap with the A beams (see Fig.~\ref{fig:gwide_switchyard}). These periscopes are implemented through flat mirrors hanging upside down from two bridges, which are motorized and integrated into VLTI's ARAL \citep{2004SPIE.5491.1079M} system for automated removal if not in use. The mirrors all ensure a peak-to-valley wavefront error of better than $\lambda/20$ ($\lambda$ = 632.8\,nm) within the footprint of the beam and were coated simultaneously to minimize differential polarisation between the beams.

The main optical delay lines do not compensate for the differential optical path length (OPD) changes that result from the up to several 10 arcsec wide on-sky separation of the A and B beams. To correct for this, PRIMA \citep{2008NewAR..52..199D} originally introduced the differential delay lines (DDLs) \citep{2005ASPC..338..167L}. After the discontinuation of the PRIMA project the DDLs were turned off. We reactivated the DDLs for the use in GRAVITY Wide. This however required a modification to relay the pupil at the correct distance for the pickup by GRAVITY. The DDLs (see Fig.~\ref{fig:gwide_ddl}) consist of a three-mirror cat's eye system with five optical reflections (it's primary and secondary mirrors are passed twice). The tertiary mirror is located in the focus of the system and its radius of curvature directly controls the distance of the output pupil plane. We replaced the M3s with new mirrors with appropriately modified radii of curvature.
The differential delay between the SC and FT is stabilized on the internal laser metrology of the PRIMA DDLs. The current GRAVITY Wide implementation does not propagate the GRAVITY laser metrology up to the telescope, and therefore does not yet provide the absolute phase and astrometry between the SC and FT. The upgrade with a full optical path length coverage by the GRAVITY metrology is foreseen for the next project phase.

With two beams entering GRAVITY, two sets of pupil beacons would appear in the acquisition camera pupil tracking images. Hence, in April 2022 we have installed a narrow band filter blocking the pupil beacon light from the A beam, such that only one set of pupil beacons appears in the acquisition camera from which to measure the lateral and longitudinal offsets. These offsets can then be used to adjust the pupil for both beams A and B. Finally, the VLTI switchyard itself required a reconfiguration to allow for this new beam routing. For this four of eight motorized flat mirrors were re-positioned.

All these modifications were carried out in December 2021 and conclude the hardware part of the first phase of the GRAVITY Wide implementation. We have begun to work on the second phase as part of the GRAVITY+ project which will remove the PRIMA DDLs entirely and save five optical reflections for the B beams. For this, in the second phase we will motorize the beam compressors of the VLTI (see again Fig.~\ref{fig:gwide_switchyard}) to take over the differential optical path length compensation. The updated mechanics will ensure that the original optical specifications of the DDL system are met.

\subsection{Software changes} 
\label{subsec:wide_software}
Along with the instrumental changes, a number of software changes had to be made to bring GRAVITY Wide into operation. One general modification is to use the VLTI field selector mirrors (FSMs) and variable curvature mirrors (VCMs) as the actuators in all control loops instead of GRAVITY's internal mirrors. In particular, the field tracking and fringe tracker beam optimization loops are now controlled through FSM A and lateral pupil tracking is controlled through VCMs A and B. Further, during acquisition of the FT and SC targets only FSMs A and B are moved to properly align them with the fibres.

We additionally developed and implemented a new ``SC Tracking'' control loop which is similar to the current field tracking loop. Here, we measure the SC target position through Gaussian fitting directly from the acquisition camera field images. The separation between the target position and SC fibre position is calculated and translated into the necessary FSM B offsets to bring the SC target back to the fibre. To help facilitate this for faint SC targets ($m_H<17$), we added the option for a longer detector integration time (DIT) (2.8s) on the acquisition camera such that the SC target would be reliably detected on the field images.

\subsection{Metrology OFF mode}
\label{subsec:gravity_faint}
\begin{figure}[t]
\centering
\includegraphics[width=0.48\textwidth]{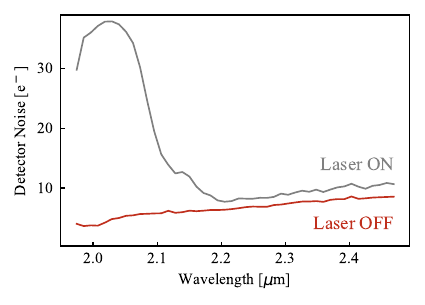}
\caption{Detector noise of GRAVITY for a sky frame with an integration time of \SI{30}{\second}. The noise is shown with the metrology laser on (in grey) and the laser turned off (in red).} \label{fig:gfaint}
\end{figure}
When using GRAVITY Wide to observe extremely faint targets it is important to reduce existing noise sources as much as possible. The dominant instrumental noise source in GRAVITY is the scattering of the metrology laser in the instrument. While the metrology laser wavelength lies outside the science wavelength, Raman scattering and backscattering from rare-earth elements in the optical fibers create a broad noise peak in the blue part of the K-band, as well as a constant background flux over the full detector \citep{2018SPIE10701E..1YL}. The detector noise of GRAVITY is shown in Fig.~\ref{fig:gfaint}. The direct back scattering is the dominant noise source from \SIrange{2.0}{2.1}{\micro\meter} and a diffuse background from the laser contributes to the noise above \SI{2.1}{\micro\meter}. 

The metrology system is only necessary for astrometric observations in the GRAVITY dual-beam mode. For observations where the science target is within a single beam, such as observations in GRAVITY on-axis or GRAVITY Wide, the metrology system is not needed. To avoid unnecessary noise in those observations, we developed a new instrument mode: the \textit{Metrology OFF} mode. In this mode the laser amplifier of the metrology laser is turned off during the observation. This removes the back scattering of the laser on the detector. The resulting noise for a \SI{30}{\second} sky frame is shown in Fig.~\ref{fig:gfaint}. The noise decreases by a factor of eight in the blue part of the spectrum (from \SIrange{2.0}{2.1}{\micro\meter}) and on average by a factor of 2.5. The remaining dominant noise sources are the read-out noise from the detector as well as the thermal background from the telescopes, beam relay and sky.

\begin{figure*}[t]
\includegraphics[width=18cm]{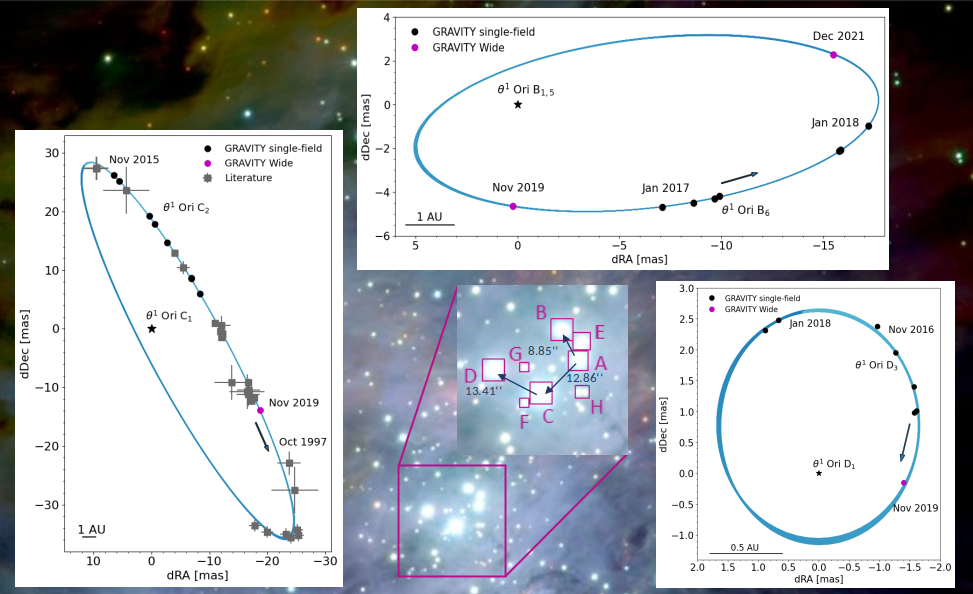}
\caption{\textit{Middle:} Orion nebula in the background and zoom on the Orion Trapezium Cluster ($\theta^1$) stars in the inset. Blue arrows mark the separation between FT and SC. \textit{Upper:} Orbit of $\theta^1$ Ori B$_6$ around the eclipsing binary $\theta^1$ Ori B$_{1,5}$ at the center. We use $\theta^1$ Ori A at a separation of 8.85 arcsec as the FT. Note that $\theta^1$ Ori B$_6$ has completed more than one orbital revolution between 2017 and 2021. \textit{Left:} Orbit of $\theta^1$ Ori C$_2$ around the primary star $\theta^1$ Ori C$_{1}$ at the center. $\theta^1$ Ori C$_2$ has completed more than two orbital revolutions. Literature data points are measurements taken from \citet{1999A&A...347L..15W}, \citet{2003A&A...402..267S}, \citet{2007A&A...466..649K}, \citet{2008ApJ...674L..97P}, \citet{2009A&A...497..195K}, and \citet{2013A&A...550A..82G}. The observation was performed with $\theta^1$ Ori A at a separation of 12.86 arcsec as the FT. \textit{Right:} Orbit of $\theta^1$ Ori D$_3$ around the primary star $\theta^1$ Ori D$_1$ at the center. We use $\theta^1$ Ori C at a separation of 13.41 arcsec as the FT. $\theta^1$ Ori D$_3$ has revolved 20 times between the first and last data point. Background image and zoom from ESO/M. McCaughrean et al. (AIP).}
\label{figure:orion_all}
\end{figure*}

\section{First GRAVITY Wide observations} 
\label{sec:observations}
\subsection{Data}
\label{subsec:data}
The first wide-angle interferometric observations were performed in five runs. The first two runs were executed between November 2019 and March 2020 with a prototype implementation of GRAVITY Wide (no proper pupil relay to GRAVITY). A third and fourth run was performed in December 2021 and January 2022, and a fifth run in April 2022. We used both the UTs and ATs. The ATs were mounted in the configuration A0-G1-J2-K0 in November 2019 and December 2021, and on the stations A0-G1-J2-J3 in March 2020. In November 2019 we carried out observations with the ATs on the Orion Trapezium Cluster. With the UTs, we performed GRAVITY Wide observations with the prototype implementation on the binary star HD 105913A in March 2020, and with the proper implementation of GRAVITY Wide on the quasars LAMOST09 in December 2021 and January 2022, and SDSS1615 in April 2022. We provide a detailed list of the observations and their parameters in Appendix \ref{appendix:data_prototype} for the prototype implementation of GRAVITY Wide, and in Appendix \ref{appendix:data_commissioning} for the data for the later runs.

In total, we observed 36 FT--SC pairs with angular separations between 2 arcsec and 32 arcsec. The observations were performed in one of the three spectral resolutions offered: low with $R = \lambda / \Delta \lambda \approx$ 20, medium with R $\approx$ 500 and high with R $\approx$ 4500. Further, the light of the FT and SC was measured in either combined or split linear polarization. The integration time on the science spectrometer was between seconds and minutes, depending on the magnitude of the science target. The data allow us to demonstrate the performance of GRAVITY Wide on the one side, and evaluate the coherence loss for increasing off-axis separations due to atmosphere anisoplanatism on the other side. Further, the FT--SC pairs include faint objects that were observed to explore the current limit of the new observing mode.
We use the GRAVITY pipeline \citep{2014SPIE.9146E..2DL, 2017A&A...602A..94G} to reduce the data.

\subsection{Binaries in the Orion Nebula}
\label{subsec:orion}
One of the main targets for science demonstration of GRAVITY Wide with the ATs is the Orion Trapezium Cluster. The cluster is located in the heart of the Orion Nebula at a distance of 414 $\pm$ 7 pc \citep{2007A&A...474..515M} from Earth. It is one of the closest regions of massive star formation \citep{1989ARA&A..27...41G, 1997AJ....113.1733H, 2007A&A...474..515M, 2008hsf1.book..483M}, and the best-studied cluster of massive stars. An interferometric study of the cluster stars has been carried out with GRAVITY in single-field mode between November 2016 and January 2018 and it has revealed that most of the massive stars are not single stars, but multiple star systems \citep{2018A&A...620A.116G}. This is expected for massive O-type stars, as they are found more often in multiple systems than low mass stars \citep{2014ApJS..215...15S}.

These systems are good targets for GRAVITY Wide for several reasons. First, the binaries are located within the field of view of GRAVITY. Additionally, many of the main components of the cluster can be used as FT targets, since they all are separated by $<$ 20 arcsec from each other. And lastly, the former GRAVITY observations provide orbits for some of the stars. This gives the unique opportunity to test and verify GRAVITY's new wide-field mode. We observed three principal components of the cluster, $\theta^1$ Ori B, $\theta^1$ Ori C, and $\theta^1$ Ori D. Compared to GRAVITY single-field mode, where one multiple system serves as both FT and SC, in GRAVITY Wide we use one multiple system as the FT and another multiple system as the SC. This is shown in the lower middle panel in Fig.~\ref{figure:orion_all}. For the observation of $\theta^1$ Ori B and $\theta^1$ Ori C, we used $\theta^1$ Ori A at a separation of 8.85 arcsec and 12.86 arcsec, respectively, as the FT. For the observation of $\theta^1$ Ori D, the FT was $\theta^1$ Ori C at a separation of 13.41 arcsec. For details about each multiple system we refer to \citet{2018A&A...620A.116G}. Here, we present the results from the observations with GRAVITY Wide, i.e. measured separation and flux ratio of the binary components, as well as orbital parameters for each of the systems, which were derived as described in Appendix~\ref{AppendixB}.

\begin{table}[t]
\caption{Binary separation and flux ratio for $\theta^1$ Ori B, $\theta^1$ Ori C and $\theta^1$ Ori D.}
\label{tab:ori_all_sep_fluxratio} 
\centering
\begin{tabular}{l | c c c}
\hline\hline
  Object & Date & Sep. [mas] & $f$ \\
 \hline
  $\theta^1$ Ori B & \text{Nov 1, 2019} & 4.65 $\pm$ 0.07 & 0.3224 $\pm$ 0.0006 \\
   & \text{Dec 15, 2021} & 15.66 $\pm$ 0.01 & 0.3504 $\pm$ 0.0022 \\
  $\theta^1$ Ori C & \text{Nov 1, 2019} & 23.39 $\pm$ 0.14 & 0.3099 $\pm$ 0.0021 \\
  $\theta^1$ Ori D & \text{Nov 2, 2019} & 1.41 $\pm$ 0.02 & 0.3419 $\pm$ 0.0064 \\
\hline
\end{tabular}
\tablefoot{
Here, given are the binary separation (sep.) and flux ratio ($f$) for the multiple star systems observed with GRAVITY Wide in November 2019 and December 2021.
}
\end{table}

\begin{table}[t]
\caption{Orbital parameters for $\theta^1$ Ori B, $\theta^1$ Ori C and $\theta^1$ Ori D.}
\label{tab:ori_all_orbital_params} 
\centering
\begin{tabular}{l | c c}
\hline\hline
    & \text{This work} & \text{(1)}  \\
\hline
\multicolumn{1}{l|}{} & \multicolumn{2}{c}{\textbf{$\theta^1$ Ori B}}\\
$a$ [mas] & 12.982 $\pm$ 0.018 & -- \\
$e$ & 0.67326 $\pm$ 0.00037 & -- \\
$i$ [$^{\circ}$] & 69.233 $\pm$ 0.063 & -- \\
$\omega$ [$^{\circ}$] & 133.48 $\pm$ 0.13 & -- \\
$\Omega$ \ [$^{\circ}$] & 282.860 $\pm$ 0.028 & -- \\
$P$ [yr] & 3.22571 $\pm$ 0.00056 & -- \\
\textbf{$t_\mathrm{P}$} [yr] & 2019.3935 $\pm$ 0.0019 & -- \\
\textbf{$M_\mathrm{tot}$} [M$_{\odot}$] & 14.928 $\pm$ 0.057 & -- \\
\hline
\multicolumn{1}{l|}{} & \multicolumn{2}{c}{\textbf{$\theta^1$ Ori C}}\\
$a$ [mas] & 44.47 $\pm$ 0.19 & 45 $\pm$ 2 \\
$e$ &  0.5908 $\pm$ 0.0034 & 0.59 $\pm$ 0.04 \\
$i$ [$^{\circ}$] &  98.702 $\pm$ 0.050 & 98.6 $\pm$ 0.6 \\
$\omega$ [$^{\circ}$] &  283.70 $\pm$ 0.14 & 283 $\pm$ 2 \\
$\Omega$ \ [$^{\circ}$] & 27.516 $\pm$ 0.055 & 27.9 $\pm$ 0.7 \\
$P$ [yr] & 11.4426 $\pm$ 0.0097 & 11.4 $\pm$ 0.2  \\
\textbf{$t_\mathrm{P}$} [yr] & 2002.304 $\pm$ 0.019 & 2002.2 $\pm$ 0.2 \\
\textbf{$M_\mathrm{tot}$} [M$_{\odot}$] & 47.70 $\pm$ 0.61 & 46.4 $\pm$ 5.9\\
\hline
\multicolumn{1}{l|}{} & \multicolumn{2}{c}{\textbf{$\theta^1$ Ori D}}\\
$a$ [mas] & 1.94 $\pm$ 0.014 & 1.86 $\pm$ 0.06  \\
$e$ & 0.391 $\pm$ 0.011 & 0.43 $\pm$ 0.03   \\
$i$ [$^{\circ}$] & 155.19 $\pm$ 0.70 & 160 $\pm$ 12  \\
$\omega$ [$^{\circ}$] & 156.40 $\pm$ 12.03 & 166 $\pm$ 27  \\
$\Omega$ \ [$^{\circ}$] & 339.56 $\pm$ 9.52 & 346 $\pm$ 24  \\
$P$ [yr] & 0.1449059 $\pm$ 0.0000086 & 0.1452 $\pm$ 0.0002  \\
$t_\mathrm{P}$ [yr] & 2017.1004 $\pm$ 0.0015 & 2017.101 $\pm$ 0.001 \\
$M_\mathrm{tot}$ [M$_{\odot}$] & 24.91 $\pm$ 0.54 & 21.68 $\pm$ 0.05 \\
\hline
\end{tabular}
\tablebib{
(1)~\citet{2018A&A...620A.116G}.
}
\tablefoot{
$a$ is the semi-major axis, $e$ the eccentricity, $i$ the inclination, $\omega$ the argument of periastron of the secondary's orbit, $\Omega$ the longitude of ascending node, $P$ the period, $t_\mathrm{P}$ the time of periastron passage, and $M_\mathrm{tot}$ the total system mass assuming a parallax of 2.415 $\pm$ 0.040 mas \citep{2007A&A...474..515M}.
}
\end{table}

We observed the multiple system $\theta^1$ Ori B in the nights of November 1st, 2019 and December 15th, 2021 with GRAVITY Wide. The measured separation and flux ratio between $\theta^1$ Ori B$_6$ and $\theta^1$ Ori B$_{1,5}$ is given in Tab.~\ref{tab:ori_all_sep_fluxratio}. We use the measurements from GRAVITY single-field mode in 2017/18 \citep{2018A&A...620A.116G} and the new measurements from GRAVITY Wide to determine orbital parameters. Fig.~\ref{figure:orion_all} presents the new orbit. The corresponding orbital parameters are shown in Tab.~\ref{tab:ori_all_orbital_params}. We find a total system mass of M$_{B_{1,5}B_6}$ = 14.93 $\pm$ 0.06 M$_{\odot}$. This result is consistent with the sum of the masses of M$_{B_1}$ = 7 M$_{\odot}$ \citep{1999A&A...347L..15W}, M$_{B_5} \sim$ 2 M$_{\odot}$ \citep{2018A&A...620A.116G} and the upper limit of \citet{2018A&A...620A.116G}, which is M$_{B_6} \sim$ 6 M$_{\odot}$.\looseness=-2

In a second observation with GRAVITY Wide, we observed $\theta^1$ Ori C in the night of November 1st, 2019. Tab.~\ref{tab:ori_all_sep_fluxratio} provides the measured separation and flux ratio of the binary components $\theta^1$ Ori C$_1$ and $\theta^1$ Ori C$_2$. We fit the orbit of $\theta^1$ Ori C with the new GRAVITY Wide measurement and with measurements from GRAVITY single-field and the literature. The orbit is presented in Fig.~\ref{figure:orion_all}.  In Tab.~\ref{tab:ori_all_orbital_params} we compare our results with the results from \citet{2018A&A...620A.116G}. We find that they agree with each other and that our fit result supports the large mass of the binary $\theta^1$ Ori  C. Further, we obtain more constrained values for the orbital parameters, and a slightly larger total system mass, which is however within the error bars of \citet{2018A&A...620A.116G}.

Finally, we observed $\theta^1$ Ori D in the night of November 2nd, 2019 with GRAVITY Wide. We give the measured separation and flux ratio between $\theta^1$ Ori D$_1$ and $\theta^1$ Ori D$_3$ in Tab.~\ref{tab:ori_all_sep_fluxratio}. We use the measurements from GRAVITY single-field \citep{2018A&A...620A.116G} and the new measurement with GRAVITY Wide and fit the orbit of $\theta^1$ Ori D$_3$ around the primary $\theta^1$ Ori D$_1$. We present the orbit in Fig.~\ref{figure:orion_all}. We compare the fit results with the results from \citet{2018A&A...620A.116G} in Tab.~\ref{tab:ori_all_orbital_params}. We notice that the uncertainties are smaller for $a$, $e$, $i$, $\omega$ and $\Omega$. Additionally, we find a larger total system mass.

The above presented results improve the results of previous observations of multiple star systems in the Trapezium. This verifies the functionality of GRAVITY Wide with the ATs, and proves that the new data points are of equal quality than the GRAVITY single-field points.

\subsection{HD 105913A}
\label{subsec:HD105913A}
We highlight the first GRAVITY Wide observation with the 8 m UTs. The binary HD 105913A was observed with GRAVITY Wide in the night of March 9th, 2020. A third companion star, HD 105913B, is located at a separation of 5.11 arcsec from HD 105913A \citep{2018A&A...616A...1G} and was used as the FT. The  binary components in HD 105913A, Aa and Ab, have a period of 211.59 days and a mass ratio of $q$ = 0.874. The triple system HD 105913 Aa,Ab,B is of spectral type K1 and located at 34 pc from the Sun \citep{2019AJ....158..222T}.

\begin{figure}[t]
\centering
\begin{subfigure}{.45\textwidth}
  \centering
  \includegraphics[width=\textwidth]{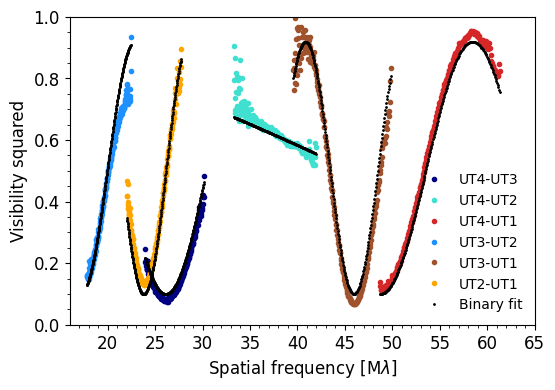}
\end{subfigure}
\begin{subfigure}{.45\textwidth}
  \centering
  \includegraphics[width=\textwidth]{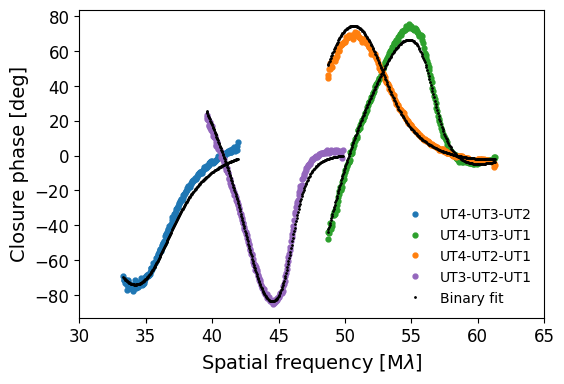}
\end{subfigure}
\caption{\textit{Upper:} Observed (color) and fitted (black) visibility squared, and \textit{Lower:} observed (color) and fitted (black) closure phases for the observation of HD 105913A on March 9th, 2020. The binary system Aa,Ab was observed with the UTs. The separation to the fringe tracker, HD 105913B, is 5.11 arcsec.}
\label{fig:hd105913ab_vis2_t3phi}
\end{figure}

\begin{figure}[ht]
\centering
\includegraphics[width=0.4\textwidth]{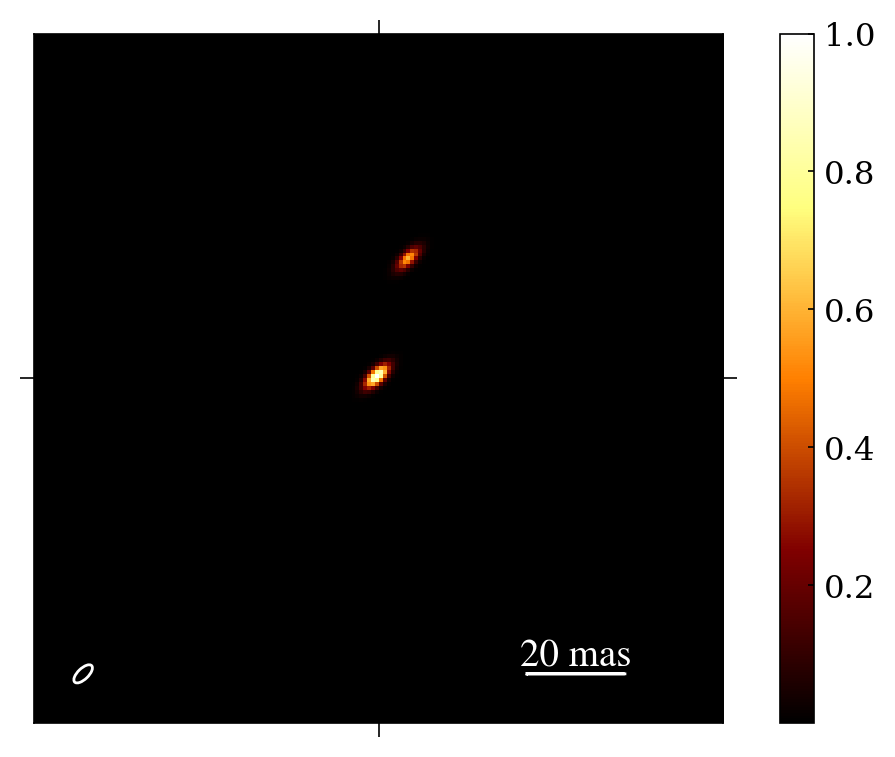}
\caption{Image of the binary HD 105913A, reconstructed with the imaging code G$^R$ \citep{2022A&A...657A..82G}. We indicate the FWHM of the corresponding dirty beam in the bottom left corner. The flux is normalized to the star Aa at the center.}
\label{figure:hd105913a}
\end{figure}

In Fig.~\ref{fig:hd105913ab_vis2_t3phi} we show the observed and fitted visibility squared and closure phases, respectively. We find the position of the secondary star at (dRA, dDec) = (-5.83 $\pm$ 0.002, 24.11 $\pm$ 0.002) mas with respect to the primary star fixed at the center, therefore the measured binary separation is 24.8 $\pm$ 0.002 mas. The error on the position, thus separation, is statistical and does not include systematics. We observe a flux ratio $f$ = 0.64 $\pm$ 0.039, thus Aa at the center is the brighter star. We show an image of the binary reconstructed with the imaging code G$^R$ \citep{2022A&A...657A..82G} in Fig.~\ref{figure:hd105913a}, which is obtained from imaging with closure phases and visibility amplitudes.

\subsection{Broad line region of redshift two quasars}
\label{subsec:qso_results}
GRAVITY has spatially resolved the broad-line region of low-redshift AGNs \citep{2018Natur.563..657G, 2020A&A...643A.154G, 2021A&A...648A.117G}.  This is achieved by measuring the differential phase of the BLR referenced to the continuum emission from the hot dust closely surrounding the BLR. Before GRAVITY Wide, however, only the brightest and therefore nearby AGN could be observed with GRAVITY because it is nearly impossible to find a bright FT star within 2 arcsec of extragalactic objects. Therefore, only on-axis GRAVITY observations were possible which imposed a $m_K < 10.5$ limit to enable fringe tracking on the AGN.

With GRAVITY Wide, we are able to observe fainter quasars at higher redshift given the increased sky coverage and ability to find off-axis FT stars.  In particular, quasars at $z \approx 2-3$  are ideal targets, as the bright H$\alpha$ line is redshifted into the $K$ band. This line is very strong compared to the continuum which boosts the observed differential phase by a factor of $\gtrsim 10$ compared to the $z \lesssim 0.1$ AGN. $z \approx 2$ is often called ``cosmic noon'', the epoch in galaxy evolution when both star formation and SMBH accretion peaked \citep{2014ARA&A..52..415M}. Since SMBHs and their host galaxies are thought to co-evolve together \citep[e.g.][]{Heckman:2014lr}, measuring SMBH masses throughout the history of the Universe and especially at cosmic noon is critical for understanding galaxy evolution.

\begin{figure*}
\includegraphics[width=\textwidth]{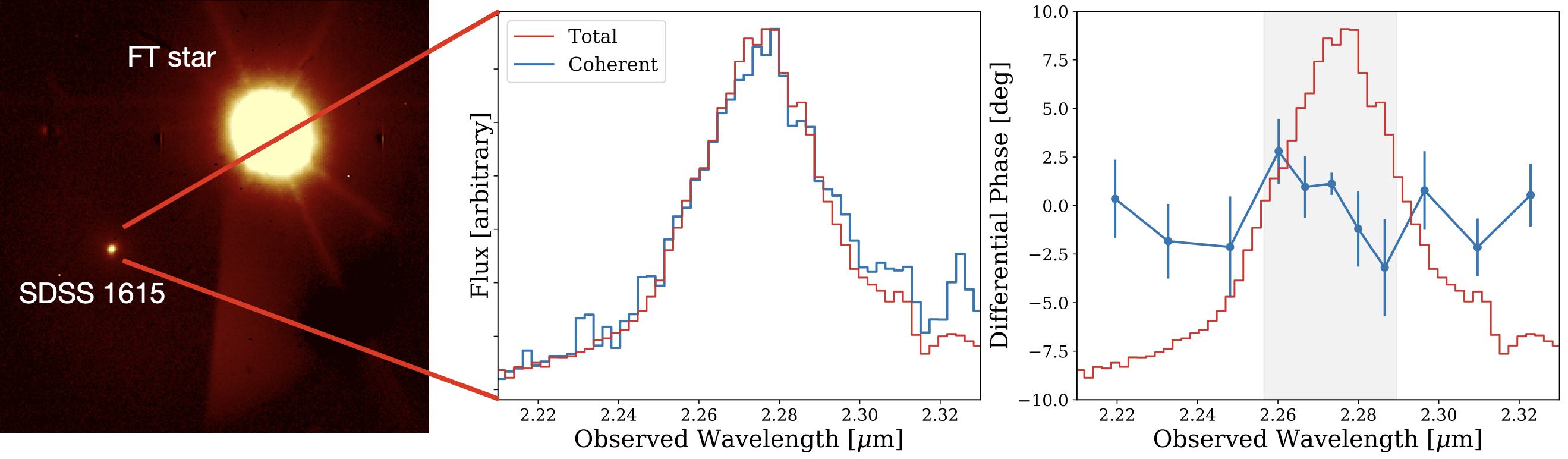}
\caption{\textit{Left:} Acquisition camera field image during observations of SDSS1615, showing successful acquisition and detection of the quasar. \textit{Middle:} Average total flux spectrum (red) over all four telescopes showing the detection of the broad H$\alpha$ line at 2.28 $\mu$m consistent with a redshift of $z=2.46$. In blue is the average coherent flux spectrum using the UT4-UT1 and UT3-UT1 baseline and the post-processing explained in the text. Coherent flux is significantly detected across the entire spectrum. \textit{Right:} Average differential phase spectrum using the UT4-UT1 and UT3-UT1 baselines overlaid on the normalized total flux spectrum. With an RMS of $\approx1-2$\degree within the FWHM of the line (gray shaded region), we tentatively detect the rotating disk signal of the BLR.}
\label{fig:qso}
\end{figure*}

We selected two quasars, LAMOST09 ($z=2.33$, m$_K=15.1$) and SDSS1615 ($z=2.46$, m$_K=15.6$), to be observed with the UT array and GRAVITY Wide. LAMOST09 has a FT star at a separation of 12.8 arcsec with m$_K=10.4$ while SDSS1615's FT star is 7.5 arcsec away with m$_K=10.45$. Observations took place on 18 December 2021, 19 January 2022, and 25 January 2022 for LAMOST09 and 17 April 2022 for SDSS1615 in MEDIUM spectral resolution. Atmospheric conditions ranged from poor to moderate over the three nights for LAMOST09 resulting in difficulty maintaining fringe tracking especially on 18 December. Still, each night we were able to acquire, fringe-track, and detect fringes with the QSO-star pair resulting in 96 min of useable data. For SDSS1615, we had very good weather conditions for the single night of observing and were able to consistently fringe-track and integrate on the QSO over 1 hour. The left panel of Fig.~\ref{fig:qso} shows an example acquisition camera field image containing both the FT star and SDSS1615 demonstrating successful acquisition.\looseness=-2

The raw data were reduced through the GRAVITY pipeline to produce complex visibilities for every DIT (60s for LAMOST09 and 100s for SDSS1615). Due to the low flux level, significant fringe jumps, and a slow drift in the OPD, we chose not to use the standard coherent integration within the pipeline and instead manually ran our own post-processing. This involved first running a 2D Discrete Fourier Transform on each individual night's dataset to determine an overall group delay accounting for the OPD drifts for each baseline. The drifts were removed for each DIT individually with the self-reference method from \cite{2007A&A...464...29T,2008SPIE.7013E..1GM}, whereby the phase reference for each spectral channel is constructed from all other spectral channels. This post-processing significantly increased the coherent flux for each baseline. We note that once the cause of the drift can be identified and potentially fixed, even longer DITs will be possible to further increase the sensitivity of QSO observations.

In the middle panel of Fig.~\ref{fig:qso}, we plot the average coherent flux overlaid on the total flux in the wavelength range of the expected H$\alpha$ line for SDSS1615. The detected H$\alpha$ line averaged over all four telescopes peaks around \SI{2.28}{\micro\meter} and has a FWHM of 4300 km s$^{-1}$. The coherent flux is averaged over two of the longest baselines, UT4-UT1 and UT3-UT1 where we expect the strongest differential phase signal. We clearly detect coherent flux across the entire spectrum, which represents the first near-infrared interferometric fringes of a high redshift object. The right panel shows the average differential phase for the same two baselines. Within the FWHM of the H$\alpha$ line (gray shaded region), we measure an RMS noise of approximately 1--2\degree and observe a tentative ``S-shape'' signal indicative of a rotating BLR. SDSS1615 is a high luminosity quasar with an estimated BLR size of $\sim1.1$ pc and SMBH mass of $10^{9.9}$ M$_{\odot}$ based on the [CIV] line profile \citep{2020ApJS..249...17R}. Using these values, the differential phase peak should be between 3--6\degree depending on the inclination and position angle on-sky of the BLR which matches well the emerging signal in Fig.~\ref{fig:qso}. 

\section{The role of atmospheric seeing for wide-angle fringe tracking}
\label{sec:discussion}
\subsection{Atmospheric coherence loss}
\label{sec:atmosphere}
Similar to classical ground-based observations in the optical and near-infrared, atmospheric turbulence also plays an important role in near-infrared interferometry. Turbulence in the atmosphere of Earth leads to a blurred image of an astronomical object with a typical diameter of around 1 arcsec. The full width at half maximum (FWHM) of the blurred image is the so-called \textit{seeing}, $\epsilon$. It depends on the individual conditions at the observational site, and is given by \citep{2007MNRAS.382.1268K}
\begin{equation}\label{equ:seeing}
    \epsilon = 0.98 \frac{\lambda}{\text{r}_0} \ ,
\end{equation}
where r$_0$ is the Fried parameter, and $\lambda$ the observed wavelength.

For interferometry, especially with off-axis fringe tracking as it is done in GRAVITY Wide, atmospheric effects are crucial. In this context, the isoplanatic angle becomes an important parameter, which defines how far from the SC the FT can be without losing coherence on the SC. Angular anisoplanatism occurs when the light from two targets separated by an angle $\theta$ experiences different phase variations as it travels through different parts of the atmosphere \citep{2000plbs.conf...71Q}. The disturbed wavefront of the on-axis star can be corrected by AO, whereas the off-axis star has residual wavefront errors. Thus, the wavefront correction degrades if the science target is further away from the fringe tracking star, which leads to a loss in coherence, thus SNR. We therefore try to understand the effects of the isoplanatic angle on the observations to estimate which atmospheric conditions are best suited for observations with GRAVITY Wide.

The isoplanatic angle is given by \citep{2000plbs.conf...71Q}
\begin{equation}\label{equ:isoplanatic_angle}
    \theta_0 = \text{0.314(cos }z) \frac{r_0}{H} \ ,
\end{equation}
where $z$ is the zenith angle, and $H$ the mean effective turbulence height which can be expressed as
\begin{equation}
    H \equiv \left( \frac{\int dh \ C^2_N(h) h^{5/3}}{\int dh C^2_N(h)} \right) ^{3/5} \ ,
\end{equation}
where $C_N^2$ is the strength of refractive index fluctuations, and $h$ the height in the atmosphere.

\begin{figure}[t]
\includegraphics[width=0.45\textwidth]{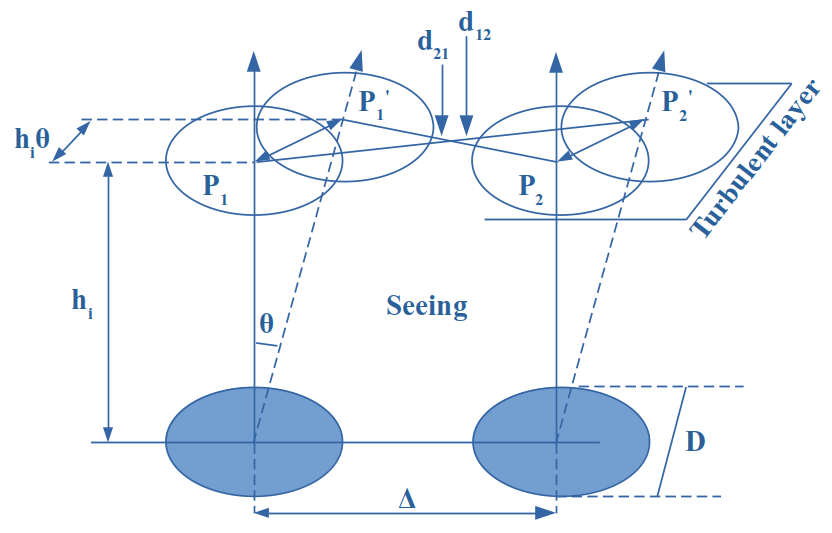}
\caption{Two-aperture interferometer with baseline length $\Delta$, and projected pupils onto a single turbulent layer at altitude $h$. The aperture diameter is $D$ and science object and phase-reference star form an angle $\theta$. Image adapted from \citet{2000A&A...353L..29E}.}
\label{fig:esposito_elements_visloss}
\end{figure}

\citet{2008A&A...477..337E} and \citet{2000A&A...353L..29E} developed a model that describes visibility reduction for off-axis fringe tracking as expected from atmosphere anisoplanatism. Fig.~\ref{fig:esposito_elements_visloss} illustrates the geometrical elements considered for the calculation. Consider a two-aperture interferometer, each with an aperture diameter $D$, with a baseline length $\Delta$. The science object and the phase-reference star are separated by $\theta$, and $P_1$, $P_2$ and $P_1^{'}$, $P_2^{'}$ are the pupils projected onto a single turbulent layer at height $h$. Further, $d_{12}$ and $d_{21}$ are the distances between the pupils $P_1$ and $P_2^{'}$, and $P_2$ and $P_1^{'}$, respectively.

Assuming that angular anisoplanatism is the only effect that reduces the visibility, the loss of visibility as a function of FT--SC separation can be approximated (following Mar\'echal) as
\begin{equation}\label{equ:vis_loss}
    V_{\text{average}}(\theta) = V \ \text{exp} \Big[ - \frac{2\pi^2}{\lambda^2}\sigma_p^2(\theta) \Big] \ ,
\end{equation}
where $V$ is the instant visibility for a delay time $\tau$ between the two optical paths, and $\sigma_p^2(\theta)$ is the anisopistonic error variance, which is the variance of the differential piston error. The expression for $\sigma_p^2(\theta)$ is derived in \citet{2008A&A...477..337E} under the following assumptions:\looseness=-2

\begin{itemize}
    \item In long-baseline interferometry, the product $h\theta$ is in general much smaller than the baseline length $\Delta$: for  $h \simeq$ 10 km, $\theta \simeq$ 15 arcsec, the product results in $h \theta \ \simeq$ 0.75 m, which is small compared to $\Delta_{UT1-UT4}$ = 130 m. Thus, the distances $d_{12}$ and $d_{21}$ can be approximated by $\Delta \simeq d_{12} \simeq d_{21}$.
    \item Further, the outer scale of turbulence $L_0$ is assumed to be infinity \citep{2015poi..book.....B}, following the Kolmogorov model. A full treatment taking into account the effect and statistics of the outer scale of turbulence can be found in \cite{2021MNRAS.506.1364B}. 
    \item \citet{2008A&A...477..337E} consider two different scenarios for deriving the isopistonic angle error. The decisive parameter for this is the fraction $\pi D/L_0$, which in our case is much smaller than one. Thus, we follow the ''small-aperture case'', which assumes that the apertures are small compared to the outer scale of turbulence $L_0$.
    \item Finally, the product ($h\theta$)$_{max}$ is equivalent to $h_{max}\theta_0$
\end{itemize}

The expression for the isopistonic angle error in Eq.~(\ref{equ:vis_loss}) is given by
\begin{equation}\label{equ:isopist_angle_error}
    \sigma_p (\theta) \sim 0.12 \pi^{1/3} \lambda \left( \frac{D}{r_0} \right) ^{-1/6} \frac{\theta}{\theta_0} \hspace{1cm}  \text{for} \ L_0 \rightarrow \infty \ .
\end{equation}
Under the assumptions described above, the anisopistonic error variance $\sigma_p (\theta)$ in Eq.~(\ref{equ:isopist_angle_error}) only depends on two geometrical elements, aperture diameter $D$ and FT--SC separation $\theta$, on two atmospheric parameters, the height of the turbulent layer $H$ and the seeing $\epsilon$, as well as on the wavelength $\lambda$ when $r_0$ and $\theta_0$ are replaced by Eq. (\ref{equ:seeing}) and Eq. (\ref{equ:isoplanatic_angle}). For observations with the VLTI the atmosphere profiler MASS-DIMM \citep{2020SPIE11448E..1KH} at the Paranal observatory measures $\theta_0$ and $\epsilon$, while $D$, $\theta$, and $\lambda$ are given by the telescopes, the FT--SC pair used, as well as the K-band wavelength observed with GRAVITY. We can therefore calculate the expected visibility of the SC for a given observation with GRAVITY Wide with Eq. (\ref{equ:isopist_angle_error}).

\subsection{Visibility as a function of FT--SC separation}
\label{subsec:vis_as_fnctn_of_sep}
\begin{figure}[t]
\includegraphics[width=0.45\textwidth]{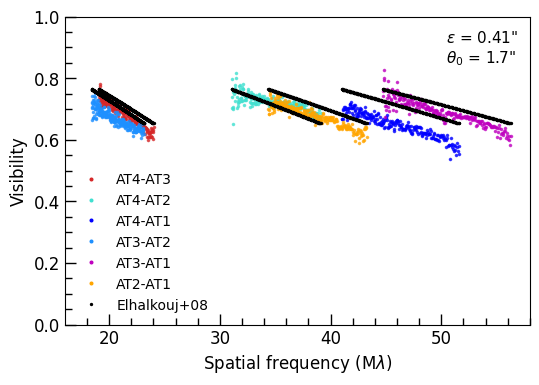}
\caption{Observed (color) and modelled (black) visibility for the star HD 48543B. The FT, HD 48543A, is located at 7.90 arcsec from the SC. During the observation, the MASS-DIMM measured a seeing $\epsilon$ of 0.41 arcsec and an isoplanatic angle $\theta_0$ of 1.7 arcsec (at 500 nm), which is 10.0 arcsec in K-band. AT1-AT2-AT3-AT4 correspond to the stations A0-G1-J2-K0. The model by \citet{2008A&A...477..337E} is able to explain the visibility loss from atmosphere anisoplanatism.}
\label{fig:HD48543A}
\end{figure}

To show how well the model for the visibilities matches the observed data we provide an example in Fig.~\ref{fig:HD48543A} for the science target HD 48543B observed with the ATs. We compare the measurement to the modelled visibilities from Eq.~(\ref{equ:vis_loss}) and Eq.~(\ref{equ:isopist_angle_error}). At the time of observation, the seeing is 0.41 arcsec (at 500 nm), and the isoplanatic angle is 1.7 arcsec (at 500 nm), which corresponds to 10.0 arcsec in K-band. We use HD 48543A at a separation of 7.90 arcsec as the FT.

\begin{figure*}[t]
\centering
\includegraphics[width=17cm]{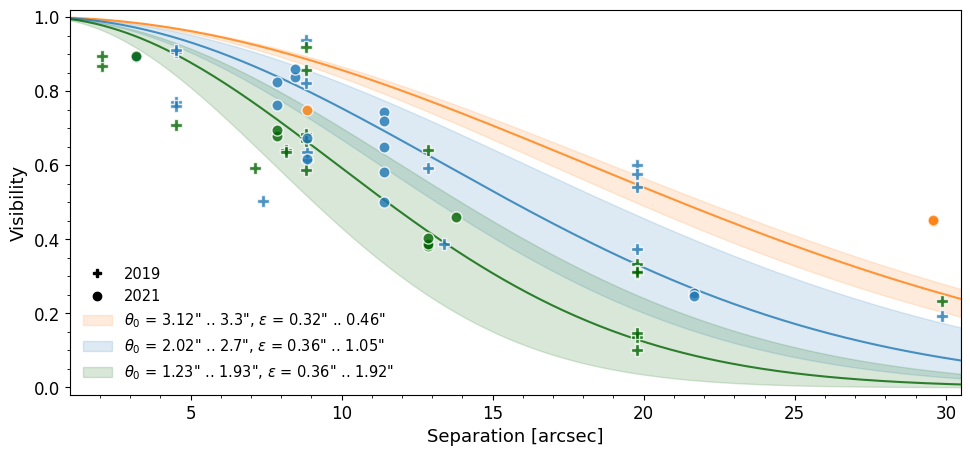}
\caption{Observed contrast loss at 2.2 $\mu$m versus off-axis separation for the ATs. The three curves indicate the typical contrast loss for different seeing and isoplanatic angle values as expected from atmospheric turbulence following \citet{2008A&A...477..337E}. Crosses represent data from GRAVITY Wide observations with the prototype implementation in November 2019, while circles represent data from GRAVITY Wide commissioning in December 2021. Seeing and isoplanatic angle values are measured by the MASS-DIMM at the Paranal observatory, and are defined at 500 nm.}
\label{fig:elhalkouj_visloss}
\end{figure*}

Fig.~\ref{fig:HD48543A} shows that the modelled visibility from atmosphere anisoplanatism matches the observed visibility very well. The visibility loss ranges from approximately 0.2 to 0.4, depending on the wavelength, in accordance with what is predicted by the model. The data presented in Fig.~\ref{fig:HD48543A} are calibrated for instrumental effects inside GRAVITY, but not for atmospheric coherence loss in the main delay lines and coherence loss of optics in the beam train from the telescopes to the VLTI lab. Therefore, we can say that the visibility loss observed in this exposure comes mostly from atmosphere anisoplanatism. Just as in Fig.~\ref{fig:HD48543A}, we compute the atmospheric visibility loss at a wavelength of 2.2 $\mu$m of 60 GRAVITY Wide AT observations. In Fig.~\ref{fig:elhalkouj_visloss}, we display each observation by either a cross (for observations from 2019) or a filled circle (for observations from 2021). Further, we sort the data in three groups based on their isoplanatic angle measured. The first group contains the highest values of $\theta_0$ with 3.12 arcsec $< \theta_0 <$ 3.3 arcsec. The second group spans 2.02 arcsec $< \theta_0 <$ 2.7 arcsec, and the third group 1.23 arcsec $< \theta_0 <$ 1.93 arcsec. Per group, we determine the mean value for both isoplanatic angle and seeing from the exposures and calculate the visibility loss with Eq.~(\ref{equ:vis_loss}) for off-axis separations up to 30 arcsec, represented by the solid orange, blue and green curve, respectively. Additionally, we color the area of the minimum and maximum visibility reduction, given by the minimum and maximum seeing and isoplanatic angle per group, respectively. 

We find that the coherence loss with increased FT--SC separation is well described by atmosphere anisoplanatism for a large outer scale of turbulence following \citet{2008A&A...477..337E}. We note two important aspects. Firstly, we find that the visibility is higher for larger isoplanatic angles. The reason for this is that a larger isoplanatic angle means a lower turbulent layer, and therefore a larger overlap of the projected pupils from the SC and FT target. This results in a better correction for wavefront aberrations by AO, as well as correction for the fringe motion of the SC. Secondly, we find that the model is more sensitive to the isoplanatic angle than to the seeing. A large isoplanatic angle is crucial for being able to observe at large off-axis separations. Until now, seeing and coherence time are taken into account for scheduling service mode observations. Based on our findings, we point out that the isoplanatic angle should be taken into account as well when executing GRAVITY Wide observations.

\subsection{Atmospheric conditions on Paranal}
\label{subsec:observing_strategy}
In Fig.~\ref{fig:elhalkouj_visloss} we see that while the overall trend is well matched by the model by \citet{2008A&A...477..337E}, only few observations match the expected visibility exactly. We discuss possible reasons in the following.

\begin{figure}[t]
\includegraphics[width=0.45\textwidth]{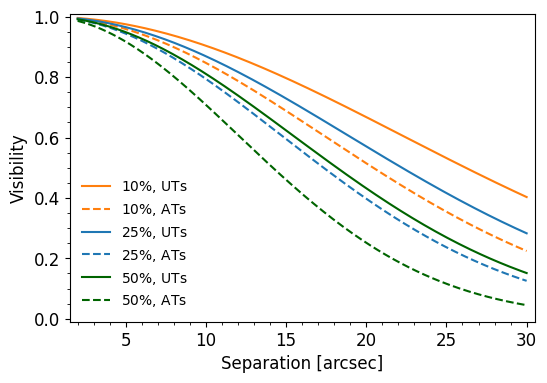}
\caption{Visibility at a wavelength of 2.2$\,\mu$m versus off-axis separation for different seeing and isoplanatic angle values given in Tab.~\ref{tab:seeing_categories}. Solid lines represent the visibility loss for the UTs, and dashed lines for the ATs, respectively.}
\label{fig:visloss_ats_uts}
\end{figure}

First, the model assumes the outer scale of turbulence to be infinity, following the Kolmogorov model. However, this scale ranges from 12 to 50 m at all major astronomical sites \citep{2016SPIE.9909E..1KZ}, and is about 22$\,$m \citep{2010Msngr.141....5M} in the atmosphere model of Paranal. This leads to an underestimation of the maximum visibility in our calculations for large telescopes  \citep{2008A&A...477..337E, 2021MNRAS.506.1364B}, but does not affect much the visibility estimates for the comparably small ATs (Fig. \ref{fig:elhalkouj_visloss}). Another point is that the model might be too simple to describe the full effects. For example, it does not take parameters such as DIT, total exposure time, coherence time, airmass or magnitude of the SC and FT into account. The target HD 10257 at 19.77 arcsec separation in Fig.~\ref{fig:elhalkouj_visloss} was observed with a sequence of DITs between 0.13$\,$s and 10$\,$s to investigate the influence of the DIT on the visibility. In the analysis we omitted the shortest DITs of 0.13$\,$s and 0.3$\,$s, because they might be short enough to "freeze" the turbulence and thus artificially increase the measured visibilities. The 60 GRAVITY Wide observations presented in Fig.~\ref{fig:elhalkouj_visloss} were performed with the ATs. For observations with the UTs we expect a higher visibility due to a larger overlap of the projected pupils on sky for given FT--SC separation. Tab.~\ref{tab:seeing_categories} presents the seeing categories for the median seeing and isoplanatic angle, respectively, for the percentiles 10$\%$, 25$\%$ and 50$\%$ measured by the MASS-DIMM at the Paranal platform. Based on these values, we compute the visibility reduction for both UTs and ATs for separations up to 30 arcsec, and present the result in Fig.~\ref{fig:visloss_ats_uts}. We can see that for the same values of seeing and $\theta_0$, the UTs provide a higher visibility.

\begin{table}[h!]
\caption{Statistics of seeing and isoplanatic angle $\theta_0$ at Paranal.}
\label{tab:seeing_categories} 
\centering
\begin{tabular}{l | c c c}
\hline\hline
   \text{Percentile}  &  \text{10$\%$} &  \text{25$\%$} &  \text{50$\%$} \\
\hline
Seeing [arcsec] at 500 nm & 0.52 & 0.62 &  0.76\\
$\theta_0$ [arcsec] at 500 nm & 3.01 & 2.48 & 1.96  \\
$\theta_0$ [arcsec] at 2.2 $\mu$m & 17.81 & 14.68 & 11.60  \\
\hline
\end{tabular}
\tablefoot{
The seeing is given at 500 nm and the isoplanatic angle $\theta_0$ at 500 nm and at 2.2 $\mu$m. Both parameters are measured by the MASS-DIMM at the Paranal observatory.
}
\end{table}

We conclude that the predictions by the model of \citet{2008A&A...477..337E} overall are in good agreement with the coherence loss we observe with GRAVITY Wide. It gives a good starting point to plan and execute observations. It also shows that it is not enough to check the seeing conditions and coherence time, but that one also needs to take the isoplanatic angle into account. Considering the performance as theoretically predicted and observationally confirmed (see Fig.~\ref{fig:HD48543A} and Fig.~\ref{fig:elhalkouj_visloss}), for GRAVITY Wide operations the offered separation will be limited to 30 arcsec, as for larger distances the coherence loss is considered too large. Fig.~\ref{fig:elhalkouj_visloss} also demonstrates that especially observations at the largest separations should be done at atmospheric conditions with a large isoplanatic angle and with a small zenith angle.

\section{Summary and outlook}
\label{sec:summary}
GRAVITY Wide has provided another breakthrough in near-infrared interferometry with the first observations using wide-angle separation fringe tracking up to about 30 arcsec across four telescopes. GRAVITY Wide significantly expands the sky coverage of GRAVITY and opens up near-infrared interferometry to new fields. In particular \textbf{we} demonstrate first near-infrared fringes of a $z=2.46$ quasar. At m$_K=15.6$, this is now the faintest extragalactic object observed by a factor of hundred along with the highest redshift. In addition, we demonstrated interferometric imaging with this new observing mode on the binary system HD 105913A, and derived new and updated orbits for several binary stars in the Orion Trapezium Cluster. 

We investigate the influence of atmospheric turbulence on the new wide-angle fringe tracking mode, and find that atmosphere anisoplanatism well describes the contrast loss in the GRAVITY Wide observations following \citet{2008A&A...477..337E}. In particular we note higher visibilities for observations with the 8 m UTs compared to the 1.8 m ATs due to a larger overlap of the projected pupils on sky. Because wide-angle separation observations are more sensitive to the isoplanatic angle than to the seeing, we propose to include the isoplanatic angle in future planning for GRAVITY Wide observations.\looseness=-2

\begin{figure}[t]
\includegraphics[width=0.45\textwidth]{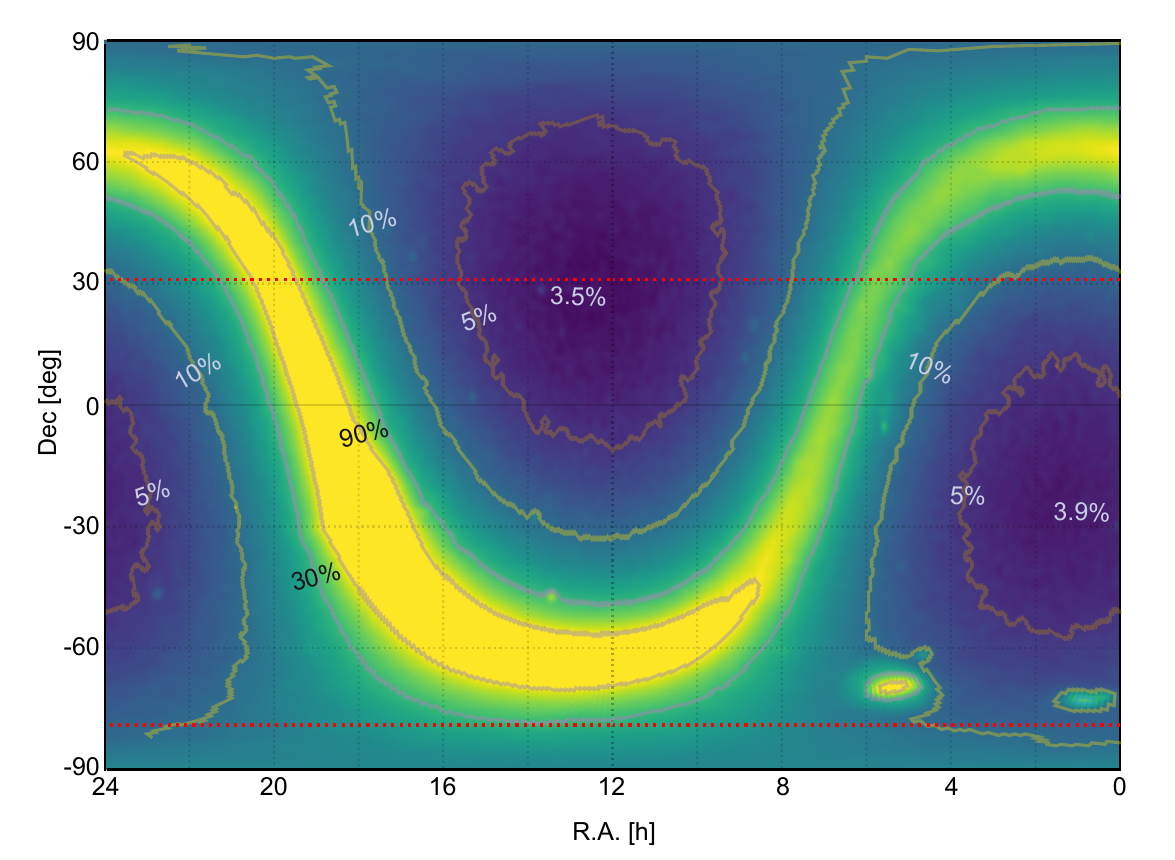}
\caption{Sky coverage for laser guide star adaptive optics supported off-axis fringe tracking with a fringe tracking star as faint as $m_K=13$, and a maximum allowed separation of 30 arcsec.}
\label{fig:sky_coverage}
\end{figure}

GRAVITY Wide is just the beginning of the full GRAVITY+ upgrade \citep{white_paper_gravityplus}. The main limitation now is the performance of the adaptive optics which has a two-fold effect in both reducing the SC target light and preventing fringe tracking on fainter stars. The next phases of GRAVITY+ therefore will install a new state-of-the-art adaptive optics system in 2024 and laser guide stars on all four UTs in 2025. The combination will allow us to push to even fainter targets across the whole sky (Fig.~\ref{fig:sky_coverage}). Together with enhanced vibration control for the telescopes and performance improvements of the GRAVITY instrument itself, we then expect fringe tracking on stars as faint as m$_K=13$ both on-axis and off-axis, and observations of objects with a magnitude up to m$_K\approx22$. The performance improvements from GRAVITY+ will open up key advances in many fields of astrophysics: e.g. the possibility to measure the spin of the Galactic Center black hole, to study SMBH growth and coevolution with galaxies over cosmic time, to directly detect exoplanets that are out of reach for traditional coronographs, to measure their atmospheric composition and orbital architecture to unprecedented precision, and to spatially resolve stars and planetary systems in formation. The leaps and bounds with regard to near-infrared interferometric AGN science can be best seen in the numbers of AGN that GRAVITY can observe at each step. GRAVITY with its original performance could observe $\approx10$ AGN at $z\sim0$ \citep{2020A&A...635A..92G}. GRAVITY Wide provides the same $10-20$ but now at $z=2$. GRAVITY+, with its full expanded capabilities, will make the jump to over 1000 AGN across cosmic time. 

The first GRAVITY Wide results presented in this paper prove the functionality of the new large separation fringe tracking mode, which will be offered to the community through ESO from October 2022 on \footnote[1]{The JMMC tech group has developed a tool for finding FT targets in order to assist in GRAVITY Wide proposal preparations. Available at \url{https://searchftt.jmmc.fr/}} and give a glimpse to the exciting future of near-infrared interferometry with GRAVITY+.

%--------------------------------------
% ACHKNOWLEDGEMENTS
%--------------------------------------
\begin{acknowledgements}
We are very grateful to our funding agencies (MPG, DFG, BMBF, ERC, CNRS (CSAA, ASHRA), Ile-de-France region (DIM ACAV+), Paris Observatory-PSL, Observatoire des Sciences de l’Univers de Grenoble, Université Grenoble Alpes, Observatoire de la Côte d'Azur, Université Côte d'Azur, and the Fundação para a Ciência e Tecnologia) and the generous support from the Max Planck Foundation - an independent, non-profit organization of private supporters of top research in the Max Planck Society. A.A. and P.G. acknowledge support from grants UIDB/00099/2020 and PTDC/FIS-AST/7002/2020. We also thank ESO and the ESO/Paranal staff, and the many scientific and technical staff members in our institutions, who helped to make GRAVITY Wide a reality. This work was supported by the European Union through ERC grant Nos. 866070 (AB, DD, RL, and MS), and has made use of the Jean-Marie Mariotti Center \texttt{Aspro} and \texttt{LITpro} services (\url{http://www.jmmc.fr/}) and of the CDS astronomical Data Centers SIMBAD and VIZIER.
\end{acknowledgements}

%--------------------------------------
% BIBLIOGRAPHY
%--------------------------------------
\bibliographystyle{aa}
\bibliography{main_gwide}
%--------------------------------------
% APPENDIX
%--------------------------------------
\onecolumn
\begin{appendix}

\section{GRAVITY Wide data}

\subsection{Data from GRAVITY Wide prototype implementation}\label{appendix:data_prototype}
We list observations performed with the prototype implementation of GRAVITY Wide, i.e. no pupil relay, in the runs in November 2019 and March 2020.
\begin{table*}[ht]
\caption{Observations with the GRAVITY Wide prototype implementation in November 2019 and March 2020.}
    \label{tab:singles_binaries_prototype}
    \begin{center}
    \begin{tabular}{l l c c c c c c c}
    \hline\hline
    SC & FT & Sep. [arcsec] & Date [UTC] & Pol. & Res. & Baseline Conf. & DIT [s] & NDIT \\
    \hline
    \noalign{\smallskip}  
    GJ 65B & GJ 65A & 2.08 & 2019 Nov 01 & S & MED & A0 G1 J2 K0 & 5.0 & 30 \\
    &  & & 2019 Nov 02 & C & MED & A0 G1 J2 K0 & 0.3 & 300 \\
    HD 105913A & HD 105913B & 5.11 & 2020 Mar 09 & S & MED & U1 U2 U3 U4 & 1.0 & 128 \\
    HD 24071 & HD 24072 & 8.16 & 2019 Nov 03 & C & MED & A0 G1 J2 K0 & 5.0 & 16 \\
    HD 218268 & HD 218269 & 8.83 & 2019 Nov 01 & S & MED & A0 G1 J2 K0 & 1.0 & 32 \\
    & & & 2019 Nov 02 & C & MED & A0 G1 J2 K0 & 0.3 & 300 \\
    & & & 2019 Nov 03 & S & MED & A0 G1 J2 K0 & 1.0 & 64 \\
    & & & 2019 Nov 03 & C & MED & A0 G1 J2 K0 & 1.0 & 64 \\
    HD 10257 & HD 10268 & 19.77 & 2019 Nov 01 & S & MED & A0 G1 J2 K0 & 5.0 & 30 \\
    & & & 2019 Nov 01 & C & MED & A0 G1 J2 K0 & 0.13 & 500 \\
    & & & 2019 Nov 02 & C & MED & A0 G1 J2 K0 & 0.3 & 300 \\
    & & & 2019 Nov 02  & C & MED & A0 G1 J2 K0 & 1.0 & 64 \\
    & & & 2019 Nov 02  & C & MED & A0 G1 J2 K0 & 2.0 & 32 \\
    & & & 2019 Nov 02  & C & MED & A0 G1 J2 K0 & 5.0 & 32 \\
    & & & 2019 Nov 02  & C & MED & A0 G1 J2 K0 & 10.0 & 32 \\
    \noalign{\smallskip}    
    \hline
    \noalign{\smallskip}
    GJ1048B & HD 16270 & 12.06 & 2019 Nov 03 & S & MED & A0 G1 J2 K0 & 10.0 & 32 \\
    LEDA 1264801 & TYC 274-754-1 & 22.14 & 2020 Mar 04  & S & MED & U1 U2 U3 U4 & 30.0 & 10 \\
    $[$VV96$]$ J095516.0  & 2MASS 09551512 & 23.93 & 2020 Mar 04  & S & MED & U1 U2 U3 U4 & 30.0 & 10 \\
    \hspace{2pt} -251732 & \hspace{2pt} -2517108 & & & & & & & \\
    QSO B0435-300 & HD 29514 & 25.99 & 2019 Nov 01  & C & MED & A0 G1 J2 K0 & 10.0 & 32 \\
    & &  & 2019 Nov 01  & C & MED & A0 G1 J2 K0 & 30.0 & 16 \\
    TYC 8071-854-1 & HD 26404 & 29.89 & 2019 Nov 02 & C & MED & A0 G1 J2 K0 & 10.0 & 30 \\
    \noalign{\smallskip}    
    \hline
    \noalign{\smallskip}
    TCC 59 & $\theta ^{1}$ Ori A & 4.23 & 2019 Nov 02 & C & MED & A0 G1 J2 K0 & 10.0 & 32 \\ 
    & & & 2019 Nov 03 & S & MED & A0 G1 J2 K0 & 10.0 & 16 \\
    $\theta ^{1}$ Ori E & $\theta ^{1}$ Ori A & 4.51 & 2019 Nov 01 & C & MED & A0 G1 J2 K0 & 5.0 & 32 \\
    & & & 2019 Nov 02 & C & MED & A0 G1 J2 K0 & 10.0 & 32 \\
    & & & 2019 Nov 03 & S & MED & A0 G1 J2 K0 & 10.0 & 16 \\
    $\theta ^{1}$ Ori F & $\theta ^{1}$ Ori C & 4.52 & 2019 Nov 02 & C & MED & A0 G1 J2 K0 & 10.0 & 32 \\
    & & & 2019 Nov 03 & S & MED & A0 G1 J2 K0 & 10.0 & 16 \\
    TCC 43 & $\theta ^{1}$ Ori B & 6.82 & 2019 Nov 02 & C & MED & A0 G1 J2 K0 & 10.0 & 32 \\ 
    $\theta ^{1}$ Ori G & $\theta ^{1}$ Ori C & 7.38 & 2019 Nov 02 & C & MED & A0 G1 J2 K0 & 10.0 & 32 \\
    $\theta ^{1}$ Ori H & $\theta ^{1}$ Ori A & 8.19 & 2019 Nov 02 & C & MED & A0 G1 J2 K0 & 10.0 & 32 \\
    $\theta ^{1}$ Ori B & $\theta ^{1}$ Ori A & 8.85 & 2019 Nov 01 & C & MED & A0 G1 J2 K0 & 5.0 & 32 \\
    & & & 2020 Mar 03 & S & MED & A0 G1 J2 J3 & 10.0 & 16 \\
    & & & 2020 Mar 03 & S & MED & A0 G1 J2 J3 & 10.0 & 32 \\
    $\theta ^{1}$ Ori C & $\theta ^{1}$ Ori A & 12.86 & 2019 Nov 01 & C & MED & A0 G1 J2 K0 & 5.0 & 32 \\
    & & & 2020 Mar 03 & S & MED & A0 G1 J2 J3 & 10.0 & 32 \\
    $\theta ^{1}$ Ori D & $\theta ^{1}$ Ori C & 13.41 & 2019 Nov 02 & C & MED & A0 G1 J2 K0 & 10.0 & 32 \\
    \noalign{\smallskip}    
    \hline
    \end{tabular}
\tablefoot{
The first block of SC--FT pairs are single and binary stars, the second block indicates a brown dwarf, three AGN, and a faint star with $m_K=9.91$ \citep{2003yCat.2246....0C} (from top to bottom). Finally, the third block lists single and binary stars within the Orion Trapezium Cluster. Columns from left to right: name of the SC, name of the FT, separation (Sep.) between SC and FT in arcsec, date of observation night in UTC, polarization (Pol.) mode in combined (C) or split (S) linear polarization, resolution (Res.) in low (LOW), medium (MED) or high (HIGH), baseline configuration (Baseline Conf.), detector integration time (DIT) in seconds and number of DITs (NDIT).
}
\end{center}
\end{table*}
\vfill

\pagebreak
\subsection{Data from GRAVITY Wide commissioning}\label{appendix:data_commissioning}
We list observations during GRAVITY Wide commissioning in December 2021 and January 2022.

\begin{table*}[ht]
    \caption{Observations from GRAVITY Wide commissioning in December 2021, January 2022 and April 2022.}
    \label{tab:singles_binaries}
    \centering
    \begin{tabular}{l l c c c c c c c}
    \hline\hline
    SC & FT & Sep. [arcsec] & Date [UTC] & Pol. & Res. & Baseline Conf. & DIT [s] & NDIT \\
    \hline
    \noalign{\smallskip}
    HD 83368A & HD 83368B & 3.28 & 2021 Dec 16 & S & MED & A0 G1 J2 K0 & 10.0 & 12 \\
     & & & 2021 Dec 16 & C & MED & A0 G1 J2 K0 & 10.0 & 12 \\
    TYC6018-535-1 & TYC6018-1742-1 & 3.91 & 2021 Dec 18 & C & HIGH & U1 U2 U3 U4 & 1.0 & 200 \\
     & & & 2021 Dec 19 & C & HIGH & U1 U2 U3 U4 & 3.0 & 30 \\
    2MASS 09325130 & IRAS 09312-5534 & 4.83 & 2021 Dec 17 & S & MED & A0 G1 J2 K0 & 30.0 & 4 \\
    \hspace{2pt} -5548206 & & & & & & & & \\
    HD 28255A & HD 28255B & 5.52 & 2021 Dec 13 & S & MED & A0 G1 J2 K0 & 10.0 & 12 \\
    SDSS J161513.84 & 2MASS	16151430 & 7.50 & 2022 Apr 17 & C & MED & U1 U2 U3 U4 & 100.0 & 6 \\
   \hspace{2pt} +084914.4 & \hspace{2pt} +0849167 & & & & & & & \\
    HD 48543B & HD 48543A & 7.90 & 2021 Dec 15 & C & MED & A0 G1 J2 K0 & 10.0 & 12 \\
     & & & 2021 Dec 15 & S & MED & A0 G1 J2 K0 & 10.0 & 12 \\
    $^*$ f Eri A & $^*$ f Eri B & 8.19 & 2021 Dec 16 & S & MED & A0 G1 J2 K0 & 10.0 & 4 \\
     & & & 2021 Dec 16 & S & MED & A0 G1 J2 K0 & 10.0 & 12 \\
     & & & 2021 Dec 16 & S & MED & A0 G1 J2 K0 & 3.0 & 12 \\
    $\theta ^1$ Ori B &  $\theta ^1$ Ori A & 8.85 & 2021 Dec 15 & S & MED & A0 G1 J2 K0 & 10.0 & 32 \\
     & & & 2021 Dec 15 & S & MED & A0 G1 J2 K0 & 30.0 & 12 \\
    HD 34750 & HD 34749 & 10.65 & 2021 Dec 14 & C & MED & A0 G1 J2 K0 & 30.0 & 4 \\
    $^*$ p Eri A & $^*$ p Eri B & 11.46 & 2021 Dec 15 & C & HIGH & A0 G1 J2 K0 & 3.0 & 32 \\
    $^*$ p Eri B & $^*$ p Eri A & 11.46 & 2021 Dec 14 & C & HIGH & A0 G1 J2 K0 & 10.0 & 16 \\
     & & & 2021 Dec 14 & S & MED & A0 G1 J2 K0 & 3.0 & 32 \\
     & & & 2021 Dec 15 & C & HIGH & A0 G1 J2 K0 & 3.0 & 32 \\
    $\theta ^1$ Ori C & $\theta ^1$ Ori A & 12.85 & 2021 Dec 14 & C & MED & A0 G1 J2 K0 & 10.0 & 32 \\
    LAMOST & Gaia 586750639 & 12.86 & 2021 Dec 18 & S & MED & U1 U2 U3 U4 & 30.0 & 12 \\
    \hspace{2pt} J092034.16 +065 & \hspace{2pt} 2466652 & &  2022 Jan 19  & S & MED & U1 U2 U3 U4 & 30.0 & 12\\
     & &  & 2022 Jan 25 & S & MED & U1 U2 U3 U4 & 30.0 & 12 \\
    2MASS 05165352 & HD 34708 & 13.80 & 2021 Dec 14 & S & MED & A0 G1 J2 K0 & 30.0 & 8 \\
    \hspace{2pt} -4811412 019 & & & & & & & & \\
     & & & 2021 Dec 16 & S & MED & A0 G1 J2 K0 & 30.0 & 12 \\
     & & & 2021 Dec 16 & S & MED & A0 G1 J2 K0 & 3.0 & 4 \\
    $^*$ gam01 Vol & $^*$ gam02 Vol & 14.14 & 2021 Dec 13 & S & HIGH & A0 G1 J2 K0 & 10.0 & 12 \\
    HD 10268 & HD 10257 & 19.77 & 2021 Dec 14 & S & MED & A0 G1 J2 K0 & 10.0 & 12 \\
    RMC 141 & W61 7-8 & 21.58 & 2021 Dec 15 & C & MED & A0 G1 J2 K0 & 30.0 & 12\\
    PKS0435-300 & HD 29514 & 25.99 & 2021 Dec 14 & S & MED & A0 G1 J2 K0 & 30.0 & 12 \\
     & & & 2021 Dec 15 & C & MED & A0 G1 J2 K0 & 30.0 & 12 \\
    2MASS J07300800 & HD 59867 & 29.11 & 2021 Dec 15 & S & MED & A0 G1 J2 K0 & 3.0 & 32 \\
    \hspace{2pt} -3939086 & & & & & & & & \\
    HD 42092 & HD 42111 & 29.17 & 2021 Dec 15 & S & MED & A0 G1 J2 K0 & 10.0 & 12 \\
    BAT99 113 & W61 7-8 & 31.92 & 2021 Dec 15 & S & LOW & A0 G1 J2 K0 & 30.0 & 12 \\
     & & & 2021 Dec 15 & C & MED & A0 G1 J2 K0 & 30.0 & 12 \\
    \hline
    \end{tabular}
\tablefoot{
Columns from left to right: name of the SC, name of the FT, separation (Sep.) between SC and FT in arcsec, date of observation night in UTC, polarization (Pol.) mode in combined (C) or split (S) linear polarization, resolution (Res.) in low (LOW), medium (MED) or high (HIGH), baseline configuration (Baseline Conf.), detector integration time (DIT) in seconds and number of DITs (NDIT).
}
\end{table*}

\twocolumn
\pagebreak
\FloatBarrier
\section{Analysis of binary star systems}\label{AppendixB}
The analysis of binary star systems in Sect. \ref{subsec:orion} proceeds in two steps. First, we use LITpro \citep{2007NewAR..51..697T} to fit a binary model to our data. In the second step, we use \texttt{orbitize!} \citep{2020AJ....159...89B} for $\theta^1$ Ori B and $\theta^1$ Ori C, and the fitting code from \citet{2009ApJ...692.1075G} for $\theta^1$ Ori D to determine orbital parameters from our separation measurements and measurements from the literature.\looseness=-2

\subsection{Binary Fitting with LITpro}\label{subsec:litpro}
For fitting a binary model to our data, we use the model fitting software LITpro which is based on the modified Levenberg–Marquardt Algorithm \citep{1983nmuo.book.....D} and includes a Trust Region Method \citep{litpro_trust_region}. It was developed by the Jean-Marie Mariotti Center (JMMC) \footnote[2]{\label{2}Available at \url{https://www.jmmc.fr/english/tools/data-analysis/litpro/}}, the french center for optical interferometry. LITpro uses the same OI Exchange format as the GRAVITY pipeline, thus we can directly upload our data and fit observables, such as squared visibilities, closure phases or visibility amplitudes.

We can choose the types of data to be fitted and take the interferometric observables visibility squared, visibility amplitude and/or closure phase. The visibility squared for a binary model is \looseness=-2
\begin{equation}\label{equ:binary_litpro}
\begin{split}
    \nu_{\text{binary}}^2 
    &= \nu_{\text{binary}}^{*} \nu_{\text{binary}} \\
    &= \frac{\nu_a^2 + f^2\nu_b^2 + 2f|\nu_a||\nu_b|\text{cos}(2\pi (u\Delta \alpha + v\Delta \beta))}{(1+f)^2} \\
    &= \frac{\nu_a^2 + f^2 \nu_b^2 + 2f|\nu_a||\nu_b|\text{cos}(2\pi \textbf{B} \cdot \textbf{s}_{\text{binary}} / \lambda)}{(1 + f)^2} \ ,
\end{split}
\end{equation}
where $\nu_{a}$ and $\nu_{b}$ are the primary star and companion star visibility, respectively, $\lambda$, is the wavelength, $f = f_{b}/f_{a}$, is the flux ratio between both stars, \textbf{B}, is the baseline and \textbf{s}$_{\text{binary}}$ $\equiv (\Delta \alpha, \Delta \beta)$ is the position of the companion star with respect to the primary star in dRA and dDEC.

The closure phase is determined by taking the argument from the \textit{bispectrum} 
\begin{equation}\label{equ:bispectrum}
    \begin{split}
    B_{123} &= \nu_{12}^{measured} \nu_{23}^{measured} \nu_{31}^{measured} \\
    &= |F_1| \ |F_2| e^{i(\Phi_1 - \Phi_2)} \nu_{12}^{true} \cdot |F_2| \ | F_3|
    e^{i(\Phi_2 - \Phi_3)} \nu_{23}^{true} \cdot \\
    &\indent |F_3| \ |F_1| e^{i(\Phi_3 - \Phi_1)} \nu_{31}^{true} \\
    &= |F_1|^2 \ |F_2|^2 \ |F_3|^2 \ \nu_{12}^{true} \cdot \nu_{23}^{true} \cdot \nu_{31}^{true} \ ,
    \end{split}
\end{equation}
between three telescopes 1,2 and 3:
\begin{equation}\label{equ:closure_litpro}
    \phi_{1-2-3} = \text{arg}(\nu_{\text{binary,1-2}} \ \nu_{\text{binary,2-3}} \ \nu_{\text{binary,3-1}}) \ ,
\end{equation}
where $\nu_{\text{binary}}$ is computed with Eq.~(\ref{equ:binary_litpro}).

From Eq.~(\ref{equ:binary_litpro}) and (\ref{equ:closure_litpro}), the parameters $\Delta \alpha$ and $\Delta \beta$ give the distance from the primary star located at (dRA, dDec) = (0, 0) mas to the companion star, and are together with the flux ratio, $f$, and a background, $f_{\text{background}}$, the parameters to be fitted. The latter variable represents the effect of visibility loss from atmospheric turbulence. We give a starting value for the companion star position, and let LITpro's fitting engine iteratively find the minimum $\chi^2$ value. At the end of the fitting procedure, we obtain the final $\chi^2$, values and standard deviations for the fitted parameters, and covariance and correlation matrices for outlining possible parameter degeneracies.

\subsection{Orbit Modelling with \texttt{orbitize!}}\label{subsec:orbitize}
\texttt{orbitize!} is an open-source Python package for orbit fitting of directly imaged exoplanets, which can also be applied to binary star systems. \texttt{orbitize!} is comprised of a parallel-tempered Affine-invariant Markov Chain Monte Carlo (MCMC) algorithm \citep{2016MNRAS.455.1919V, 2013PASP..125..306F}, and is based on Bayesian statistics. In \texttt{orbitize!} the input data used in the orbit-fitting includes the date of the observation in MJD, and the dRA and dDec offsets from the primary star at (0,0) mas with corresponding uncertainties of the secondary star in mas. The posterior distribution over the orbital parameters, are then computed using Bayes' theorem.

The MCMC algorithm in \texttt{orbitize!} represents an efficient sampling method to explore the likelihood over a multidimensional parameter space and to determine expectation values of the model parameters, i.e. the orbital parameters. Further, it is possible to use several Markov chains, called \textit{walkers}, in parallel to fully explore the parameter space at different ''temperatures''. A higher temperature helps the walkers not to become stuck in regions of local minima. A larger number of walkers increases the samples used, but also increases the computation time.

We use 20 temperatures, 1000 walkers, and 1 million steps to ensure convergence. Further, we impose a Gaussian prior on the measurement for the mass, and we use the parallax of 2.415 $\pm$ 0.040 mas from \citet{2007A&A...474..515M}, which is fixed in our analysis. If we have an initial guess for best-fit orbital parameters, for example from the literature, we use lower and upper bounds of a uniform prior around this value. This was done for the orbit of $\theta^1$ Ori C (see Sect.~\ref{subsec:orion}), where we set bounds around orbital parameters that were already determined in the literature \citep{2018A&A...620A.116G}. In the case of $\theta^1$ Ori B (see Sect.~\ref{subsec:orion}), there are no previously determined orbital parameters, and therefore we let \texttt{orbitize!} find the best-fit orbit without giving an initial guess.
%--------------------------------------
\end{appendix}
\end{document}